\documentclass[a4paper,11pt]{article}
\usepackage{tabularx} 
\usepackage{amsmath}  
\usepackage{amssymb}  
\usepackage{graphicx} 
\usepackage[margin=1in,letterpaper]{geometry} 
\usepackage[final]{hyperref} 
\hypersetup{
	colorlinks=false,       
	linkcolor=blue,        
	citecolor=blue,        
	filecolor=magenta,     
	urlcolor=blue         
}

\usepackage[affil-it]{authblk}  

\usepackage[font=footnotesize,labelfont=bf]{caption} 

\usepackage{physics} 
\usepackage{slashed} 
\usepackage{mathtools}

\usepackage{xcolor} 
\usepackage{xfrac}

\usepackage{float} 
\usepackage{multirow}
\usepackage{makecell}
\usepackage{subcaption}
\usepackage{wrapfig}
\usepackage{enumitem}

\usepackage{bbold}
\usepackage{bm}
\usepackage[normalem]{ulem}
\usepackage{mathrsfs}

\usepackage{titlesec}
\titleformat*{\subsection}{\bfseries\boldmath}
\titleformat*{\paragraph}{\bfseries\boldmath}

\usepackage{siunitx}
\DeclareSIUnit\ecm{\textit{e} \;\text{cm}}

\renewcommand{\L}{\mathcal{L}}

\renewcommand{\O}{\mathcal{O}}

\newcommand{\PL}{\text{P}_\text{L}}
\newcommand{\PR}{\text{P}_\text{R}}

\newcommand{\lambar}{\overline\lambda}

\newcommand{\SMref}{Aoyama:2012wk,Volkov:2019phy,Volkov:2024yzc,Aoyama:2024aly,Parker:2018vye,
	Morel:2020dww,Fan:2022eto,Czarnecki:2002nt,Gnendiger:2013pva,Ludtke:2024ase,Hoferichter:2025yih,
	RBC:2018dos,Giusti:2019xct,Borsanyi:2020mff,Lehner:2020crt,Wang:2022lkq,Aubin:2022hgm,Ce:2022kxy,
	ExtendedTwistedMass:2022jpw,RBC:2023pvn,Kuberski:2024bcj,Boccaletti:2024guq,Spiegel:2024dec,
	RBC:2024fic,Djukanovic:2024cmq,ExtendedTwistedMass:2024nyi,MILC:2024ryz,Bazavov:2024eou,
	Keshavarzi:2019abf,DiLuzio:2024sps,Kurz:2014wya,Colangelo:2015ama,Masjuan:2017tvw,Colangelo:2017fiz,
	Hoferichter:2018kwz,Eichmann:2019tjk,Bijnens:2019ghy,Leutgeb:2019gbz,Cappiello:2019hwh,
	Masjuan:2020jsf,Bijnens:2020xnl,Bijnens:2021jqo,Danilkin:2021icn,Stamen:2022uqh,Leutgeb:2022lqw,
	Hoferichter:2023tgp,Hoferichter:2024fsj,Estrada:2024cfy,Deineka:2024mzt,Eichmann:2024glq,
	Bijnens:2024jgh,Hoferichter:2024bae,Holz:2024diw,Cappiello:2025fyf,Colangelo:2014qya,Blum:2019ugy,
	Chao:2021tvp,Chao:2022xzg,Blum:2023vlm,Fodor:2024jyn}

\newcommand{\expref}{Muong-2:2023cdq,Muong-2:2024hpx,Muong-2:2021ojo,
	Muong-2:2021vma,Muong-2:2021ovs,Muong-2:2021xzz,Muong-2:2006rrc}

\usepackage[style=numeric-comp, backend=bibtex, sorting=none]{biblatex}
\title{\bfseries\boldmath\Large From Higgs physics to lepton flavour violation: \\current bounds and future prospects for vector-like lepton models}

\author[1]{Gregor Daberstiel \thanks{gregor.daberstiel@mailbox.tu-dresden.de}}
\author[1]{Kilian Möhling \thanks{kilian.moehling@tu-dresden.de}}
\author[1]{Dominik Stöckinger \thanks{dominik.stoeckinger@tu-dresden.de}} 
\author[1]{\\Hyejung Stöckinger-Kim \thanks{hyejung.stöckinger-kim@tu-dresden.de}} 
\affil[1]{\small Institut für Kern- und Teilchenphysik, TU Dresden, Zellescher Weg 19, 01069 Dresden, Germany}
\date{}

\begin{document}
\maketitle


\begin{abstract}
	\noindent We present a comprehensive phenomenological study of a class of six vector-like lepton models with seesaw-like mass contributions and strong modifications to Higgs and lepton phenomenology. We focus on lepton flavour conserving and violating observables across all lepton generations and collider observables like $Z$ and Higgs decays. 
	In light of the recent progress at the LHC and precision measurements such as muon $g-2$,
	as well as in anticipation of upcoming experiments like MEGII, Mu2e/COMET, Mu3e, Belle II and at the HL-LHC, 
	we systematically survey the viable parameter space and identify
	patterns and correlations.
	The considered class of models gives rise to a rich and testable phenomenology 
	with robust and complementary probes that allow to distinguish between models in the coming experimental era. 
\end{abstract}

\setcounter{tocdepth}{2}
\tableofcontents

\section{Introduction}
Some of the most intriguing open questions in particle physics are
related to the
origin of mass and flavour of fundamental particles. In the Standard
Model (SM) the fermion masses arise after electroweak symmetry breaking (EWSB) via
Yukawa couplings to the Higgs field. 
At the same time, these Yukawa couplings define the flavour
structure and provide a source of CP violation described by the phase in the CKM matrix. However, the SM Higgs and Yukawa sectors do not explain the large fermion mass hierarchies and small CKM mixing angles,
and the amount of CP violation is not sufficient  for a dynamical
generation of the matter--antimatter asymmetry in the universe.

A large and popular class of beyond-the-SM (BSM) scenarios motivated
by these issues are models with extended Higgs sectors
\cite{LHCHiggsCrossSectionWorkingGroup:2016ypw}. An alternative
direction is to extend the SM fermion sector.
In the present paper we focus on a class of models with additional fermions that strongly alter the structure of Yukawa couplings and consequently the
resulting mass-generation mechanism as well as the Higgs interactions
of leptons. The models contain a pair of additional vector-like leptons
(VLL) with different SU$(2)_L$ quantum numbers and a rich BSM Yukawa sector. 

The VLL quantum numbers allow fundamental Dirac mass terms and Yukawa
couplings between SM leptons and VLLs  \cite{delAguila:2008pw}  as
well as Yukawa couplings between two different VLLs. The
six possible models of this kind, where the mixing between ordinary and vector-like leptons leads to seesaw-like contributions to lepton masses and chirally enhanced contributions to many observables, were classified in Ref.~\cite{Kannike:2011ng}.
While we follow Ref.~\cite{Kannike:2011ng} and focus on the lepton sector here, similar extensions in the quark sector can also be considered. 
For general investigations on vector-like fermions we refer to Refs.~\cite{Langacker:1988ur,delAguila:1989rq,Dimopoulos:1990kc,Fujikawa:1994we,Sher:1995tc,Thomas:1998wy,Frampton:1999xi,delAguila:2008pw,
	Ishiwata:2011hr,Kannike:2011ng,Buras:2011ph,Joglekar:2012vc,Kearney:2012zi,Dermisek:2013gta,Ishiwata:2013gma,Falkowski:2013jya,Freitas:2014pua,
	Ellis:2014dza,Ishiwata:2015cga,Poh:2017tfo,Patel:2017pct,Alonso:2018bcg,Endo:2020tkb,Crivellin:2020ebi,Bhattacharya:2021ltd,Cherchiglia:2021syq,Jana:2021tlx,Guedes:2021oqx,Guedes:2022cfy,Baspehlivan:2022qet,Dermisek:2022aec,Dermisek:2023nhe,Arkani-Hamed:2021xlp,Hamaguchi:2022byw,Mohling:2024qvk,deBlas:2025pco,
	Erdelyi:2025axy,Arkani-Hamed:2026wwy}.

In the past, striking predictions of these six VLL models for the lepton masses $m_i$,
Higgs coupling $Y^h_{e,ii}$ and the magnetic and electric dipole moments $\Delta a_i$ and $d_i$ have been
identified \cite{Kannike:2011ng,Dermisek:2013gta,Falkowski:2013jya,Hamaguchi:2022byw,Dermisek:2022aec,Dermisek:2023nhe}. 
They are summarised by the simple equations (based on the dominant contributions, see also the
discussion in Ref.~\cite{Athron:2025ets})
\begin{subequations}\label{previewVLLrelations}
	\begin{align}
		m_i & =( m_i^H + m_i^{HHH}) e^{-i\theta_i}, \label{eq:mass-correction}
		\\
		Y^h_{e,ii} &= \frac{1}{v} \left(m_i^{H}  + 3  m_i^{HHH} \right)e^{-i\theta_i} \label{eq:Yukawa-correction}
		\\
		\intertext{and}
		\Delta a_i &= - \frac{ \mathcal{Q} m_i }{16\pi^2 v^2}\Re \Big\{m_i^{HHH} e^{-i\theta_i}\Big\}, \label{eq:AMM-correction}
		\\
		d_i &= +\frac{\mathcal{Q} e}{32\pi^2 v^2} \Im \Big\{m_i^{HHH} e^{-i\theta_i}\Big\}, \label{eq:EDM-correction}
	\end{align}
\end{subequations}
where $m_i^H$ is the SM-like Yukawa term and
$m_i^{HHH}$ the new seesaw-like contribution to the lepton mass, 
$\theta_i$ is a CP violating complex phase and the model dependent coefficient $\mathcal{Q}$
is discussed in Sec.~\ref{sec:LFC} (see also Tab.~\ref{tab:cpl-coeff}).
 
Especially for the muon or the electron, the new seesaw-like
contribution could in principle be a substantial correction and may
reach $\O(100\%)$. The second Eq.~\eqref{eq:Yukawa-correction} shows the potentially huge impact on the effective
lepton--Higgs coupling compared to its SM value, resulting from the
combinatorial factor 3. The third and fourth equations \eqref{eq:AMM-correction} and \eqref{eq:EDM-correction} 
illustrate the strong correlation with VLL contributions to the lepton dipole moments. 
The structure of these correlations differs from the one in most other models where a
BSM mass suppression would appear on the r.h.s.~of
Eq.~\eqref{eq:AMM-correction} instead of the loop suppression (for general discussions see
Refs.~\cite{Athron:2025ets,Czarnecki:2001pv,Stockinger:2022ata,Fajfer:2021cxa,Crivellin:2021rbq,Dermisek:2022aec,Dermisek:2023nhe,Lindner:2016bgg,Valori:2025hlp} and references therein). 

In this way, the VLL models offer a unique connection between the fundamental properties
of the leptons and enable complementary probes from both high-energy Higgs-physics
and low-energy precision measurements.

Besides flavour conserving observables, important additional insights can be obtained
from charged lepton flavour violating (CLFV) processes such as $h\to\mu\tau$,
$\tau\to 3e$ or $\mu\to e$ conversion \cite{Goudelis:2011un,Ishiwata:2011hr,Ishiwata:2013gma,Falkowski:2013jya,Poh:2017tfo,Raby:2017igl,Crivellin:2018qmi}.
For many of the relevant observables, significant experimental advances
have been achieved in recent years and further progress is expected in upcoming measurements.

In light of this, we present a comprehensive study of all six VLL seesaw models.
We evaluate a multitude of observables ranging from Higgs decays such
as $h\to\mu\mu,\tau\tau$
and electroweak precision measurements of the lepton--$Z$ couplings to
low-energy lepton observables and derive model-specific patterns of correlations as well as
characterise currently preferred parameter regions.

We begin in Sec.~\ref{sec:Models} with a definition of the six VLL
models and useful mass-insertion approximations for all
observables. Sec.~\ref{sec:Constraints} then collects basic
constraints from non-chiral observables measured at LHC and
LEP. Sec.~\ref{sec:LFC} focuses on chirality flipping but flavour
conserving observables such as the ones in Eq.~\eqref{previewVLLrelations}. It
evaluates the predicted correlations  in the different models and
draws conclusions from the recent results especially on $\Delta a_\mu$
and on $h\to\mu\mu$.

Finally, Sec.~\ref{sec:LFV} presents an extensive
phenomenological analysis considering also
lepton-flavour violating observables both in the $\mu$--$e$ sector and
in the $\tau$--$\mu/e$ sectors, and from low-energy observables as well
as in $Z$ or Higgs decays. It analyses the role of chiral enhancements
and shows that the patterns of correlations are often very different
from the simple pattern implied by the familiar limit of dipole
dominance. It highlights typical features shared by all six VLL models
and identifies observables that can allow future experiments to
discriminate between the models.

\section{Charged Seesaw models}
\label{sec:Models}

\subsection{Model definitions}

\begin{table}[t]
	\centering
	\def\arraystretch{1.2}
	\begin{tabular}{c||c|c|c|c|c|c}
		\textbf{Rep.} & $(\bm1,-1)$ & $(\bm1, 0)$ & $(\bm2, -1/2)$ & $(\bm2, -\frac{3}{2})$ & $(\bm3, -1)$ & $(\bm3, 0)$ \\ \hline
		\textbf{Name} & $E$ & $N$ & $L$ & 
		$L_{\frac{3}{2}}$ & $E^a$ & $N^a$ \\
		& $E$ & $N$ & $\begin{pmatrix} L^0 \\ L^- \end{pmatrix}$ & $\begin{pmatrix} L^{-}_{\frac{3}{2}} \\ L^{--}_{\frac{3}{2}}	\end{pmatrix}$
		& $\begin{pmatrix}
			E^- & \sqrt{2} E^0 \\ \sqrt{2} E^{--} & -E^-
			\end{pmatrix}$ & $\begin{pmatrix}
			N^0 & \sqrt{2} N^+ \\ \sqrt{2} N^- & -N^0 \end{pmatrix}$ \\
	\end{tabular}
	\caption{Allowed representations of VLLs. The first number denotes the SU$(2)_L$ representation and the second
	the hyper-charge $Y$. All fields are singlets under SU$(3)_c$. 
	In the second row we show the components together with the corresponding electric charge $Q=T^3+Y$ (upper index).
	For the triplet we list the components of $\sigma^a E^a$ and $\sigma^a N^a$.}
	\label{tab:VLL-rep}
\end{table}

Here we define the six VLL models with seesaw-like contributions to
charged lepton masses~\cite{Kannike:2011ng}. Symbolically, the models
are
	\begin{align}\label{chargedseesawpossibilities}
		L\oplus E, \qquad L\oplus N, \qquad L_{\frac{3}{2}}\oplus E, \qquad L\oplus E^a, \qquad L\oplus N^a, \quad \text{and} \quad L_{\frac{3}{2}}\oplus E^a.
	\end{align}
Here $L$, $L_{\frac{3}{2}}$ denote doublets with hypercharge $Y=-1/2$,
$-3/2$; $E$ and $E^a$ are a singlet/triplet with $Y=-1$; and
$N$, $N^a$ are a singlet/triplet with $Y=0$, see also
Tab.~\ref{tab:VLL-rep}. All these fields are vector-like fermions and
carry one unit of lepton number. The quantum numbers allow four kinds
of Yukawa couplings in each model.
The additional lepton Yukawa interaction Lagrangians for the six
different models are thus given by
\begin{subequations}
	\begin{alignat}{5}
		&\mathcal{L}_{L\oplus E}& &\supset - \lambda^L_j (\overline{L}\Phi) e_{Rj}  & &- \lambda^E_i (\overline{l_{L}}_i\Phi) E 
		&&- (\overline{L}\Phi) \big[\lambda\PR+\lambar^*\PL\big]E &&+ h.c. ,
		\\[.3cm]
		&\mathcal{L}_{L\oplus N}& &\supset - \lambda^L_j (\overline{L}\Phi) e_{Rj}  & &- \lambda^N_i (\overline{l_{L}}_i\tilde\Phi) N 
		&&- (\overline{L}\tilde\Phi)\big[\lambda\PR+\lambar^*\PL\big]N  &&+ h.c. ,
		\\[.3cm]
		&\mathcal{L}_{L_{\frac{3}{2}}\oplus E}& &\supset - \lambda^L_j (\overline{L}_{\frac{3}{2}}\tilde\Phi) e_{Rj}  &&- \lambda^E_i (\overline{l_{L}}_i\Phi) E 
		&&- (\overline{L}_{\frac{3}{2}}\tilde\Phi)\big[\lambda\PR+\lambar^*\PL\big]E  &&+ h.c.,
		\\[.3cm]
		&\mathcal{L}_{L\oplus E^a}& &\supset - \lambda^L_j (\overline{L}\Phi) e_{Rj}  &&- \lambda^E_i (\overline{l_{L}}_i\sigma^a \Phi) E^a 
		&&- (\overline{L}\sigma^a\Phi) \big[\lambda\PR+\lambar^*\PL\big]E^a &&+ h.c. ,
		\\[.3cm]
		&\mathcal{L}_{L\oplus N^a}& &\supset - \lambda^L_j (\overline{L}\Phi) e_{Rj}  &&- \lambda^N_i (\overline{l_{L}}_i\sigma^a \tilde\Phi) N^a
		&&- (\overline{L}\sigma^a\tilde\Phi)\big[\lambda\PR+\lambar^*\PL\big]N^a  &&+ h.c. ,
		\\[.3cm]
		&\mathcal{L}_{L_{\frac{3}{2}}\oplus E^a}& &\supset - \lambda^L_j (\overline{L}_{\frac{3}{2}}\tilde\Phi) e_{Rj}  &&- \lambda^E_i (\overline{l_{L}}_i\sigma^a \Phi) E^a
		&&- (\overline{L}_{\frac{3}{2}}\sigma^a\tilde\Phi)\big[\lambda\PR+\lambar^*\PL\big]E^a  &&+ h.c.
	\end{alignat}
\end{subequations}
Each model contains Yukawa couplings of VLL to right-handed SM leptons $e_R$
($\lambda^L$), to left-handed SM leptons $l_L$ ($\lambda^E$ or $\lambda^N$)
and between the two VLLs ($\lambda$ and $\lambar$). Lower indices
$i,j=1,2,3$ here denote the lepton generation. The SM Higgs
doublet is denoted as $\Phi$, and $\tilde{\Phi}=i\sigma^2\Phi^*$.
In addition, each of the VLL models adds a Dirac mass term of the form
$\mathcal{L}_D = M_F \overline{F_L} F_R + h.c.$, where $F$ is any of
the VLL fields.

Before moving on to the mass basis it is interesting to point out some
redundancies in the definition of the Lagrangians.
In models with $L$ or $E$, additional gauge invariant Dirac mass terms of the type $\overline{l_{Li}}L_R$ or $\overline{E_L}e_{Ri}$ 
could be added which, however, can always be removed again by an appropriate field redefinition that mixes fields in the same representation
of the SM gauge group. Furthermore, this still allows for another unitary transformation of the SM-like fields to diagonalise the
SM Yukawa coupling matrix (see Ref.~\cite{Mohling:2024qvk}).
Another related redundancy appearing in all models stems from the possibility of complex couplings.
While, a priori, all Yukawa couplings and Dirac masses can be complex, some of their phases can be absorbed into the
VL and SM lepton fields. As a result, only three \emph{physical} CP violating phases exist, which could be chosen e.g.\ as the phases of
the SM couplings $y_i$, while all new BSM couplings could be assumed
real and positive. For the purposes of the present paper, however,  we find it
more convenient to keep generic complex couplings, instead of choosing
certain parameters to be real.

\subsection{Mass diagonalisation and mass eigenstates}\label{sec:mass-basis}

After electroweak symmetry breaking (EWSB) the Higgs couplings and
Dirac Mass terms lead to the following mass matrices for the singly
charged and neutral leptons 

\begingroup
\allowdisplaybreaks
\begin{subequations}\label{eq:mass-matrices}
	\begin{alignat}{4}
		&\mathscr{M}^-_{L\oplus E} &&= \bordermatrix{ & e_{Rj} & E_R & L^-_R \cr 
				\overline{e_{L}}_{i} & y_{ij} v & \lambda^E_i v & 0 \cr 
				\overline{L_L^-} & \lambda_j^L v & \lambda v &  M_L \cr
				\overline{E_L} & 0 & M_E & \bar\lambda v \cr} & \qquad
		&\mathscr{M}^0_{L\oplus E}& &= \bordermatrix{ & 0 & L^0_R \cr 
				\overline{\nu_{L}}_{i} & 0 & 0 \cr 
				\overline{L_L^0} & 0 &  M_L },
		\\[.3cm]
		&\mathscr{M}^-_{L\oplus N} &&= \bordermatrix{ & e_{Rj} & L^-_R \cr	
				\overline{e_{L}}_{i} & y_{ij} v & 0 \cr
				\overline{L_L^-} & \lambda^L_j v & M_L } & \qquad
		&\mathscr{M}^0_{L\oplus N} &&= \bordermatrix{ & 0 & L^0_R & N_R \cr 
				\overline{\nu_{L}}_{i} & 0 & 0 & \lambda^N_i v \cr 
				\overline{L_L^0} & 0 &  M_L & \lambda v \cr
				\overline{N_L} & 0 & \lambar v & M_N},
		\\[.3cm]
		&\mathscr{M}^-_{L_{\frac{3}{2}}\oplus E} &&= \bordermatrix{ & e_{Rj} & E_R & L^-_{\frac{3}{2}R} \cr	
				\overline{e_{L}}_i & y_{ij} v & \lambda^E_i& 0 \cr 
				\overline{L_{\frac{3}{2}L}^-} & \lambda^L_j v & \lambda v & M_L \cr 
				\overline{E_L} & 0 & M_E & \bar\lambda v} & \qquad
		&\mathscr{M}^0_{L_{\frac{3}{2}}\oplus E} &&= 0,
		\\[.3cm]
		&\mathscr{M}^-_{L\oplus E^a} &&= \bordermatrix{ & e_{Rj} & E^-_R & L_R^- \cr	
				\overline{e_{L}}_i & y_{ij} v & - \lambda_i^E v & 0 \cr 
				\overline{L^-_L} & \lambda^L_j v & -\lambda v & M_L \cr
				\overline{E_L^-} & 0 & M_E & -\bar\lambda v} & \qquad
		&\mathscr{M}^0_{L\oplus E^a} &&= \bordermatrix{ & 0 & L^0_R & E^0_R \cr 
				\overline{\nu_{L}}_{i} & 0 & 0 & \sqrt{2}\lambda^E_i v \cr 
				\overline{L_L^0} & 0 &  M_L & \sqrt{2}\lambda v \cr
				\overline{E^0_L} & 0 & \sqrt{2}\,\lambar v & M_E},
		\\[.3cm]
		&\mathscr{M}^-_{L\oplus N^a} &&= \bordermatrix{ & e_{Rj} & N^-_R & L_R^- \cr	
				\overline{e_{L}}_i & y_{ij} v & \sqrt{2} \lambda_i^N v & 0 \cr 
				\overline{L^-_L} & \lambda^L_j v & \sqrt2 \lambda v  & M_L \cr
				\overline{N_L^-} & 0 & M_N & \sqrt{2}\bar\lambda v} & \qquad 
		&\mathscr{M}^0_{L\oplus N^a} &&= \bordermatrix{ & 0 & L^0_R & N^0_R \cr 
				\overline{\nu_{L}}_{i} & 0 & 0 & \lambda^N_i v \cr 
				\overline{L_L^0} & 0 &  M_L & \lambda v \cr
				\overline{N^0_L} & 0 & \lambar v & M_N},
		\\[.3cm]
		&\mathscr{M}^-_{L_{\frac{3}{2}}\oplus E^a} &&= \bordermatrix{ & e_{Rj} & E^-_R & L_{\frac{3}{2}R}^- \cr	
				\overline{e_{L}}_i & y_e^{ij} v & -\lambda_E^i v & 0 \cr 
				\overline{L_{\frac{3}{2}L}^-} & \lambda_L^i v & \lambda v & M_L \cr
				\overline{E_L^-} & 0 & M_E & \bar\lambda v} & \qquad
		&\mathscr{M}^0_{L_{\frac{3}{2}}\oplus E^a} &&= \bordermatrix{ & 0 & E^0_R \cr 
				\overline{\nu_{L}}_{i} & 0 & \sqrt2\lambda^E_iv \cr 
				\overline{E_L^0} & 0 &  M_E },
	\end{alignat}
\end{subequations}
\endgroup
with the convention of a vacuum expectation value (vev) around $v\approx \SI{174}{GeV}$.
As in Ref.~\cite{Kannike:2011ng} we have included the corresponding
names of the multiplet components that mix via these mass matrices\footnote{However, we find some sign differences compared to
Ref.~\cite{Kannike:2011ng} as well as differences in the model-dependent coefficients Tab.~\ref{tab:cpl-coeff} later,
most notably $\mathcal{Q}$ which disagrees with Refs.~\cite{Kannike:2011ng,Dermisek:2022aec} but agrees with the later correction in Ref.~\cite{Dermisek:2023nhe}.}.
The exotic states $N^+, E^{--}$ and $L^{--}$ retain their Dirac mass after EWSB, except in the case of $L_{\frac{3}{2}}\oplus E^a$
which contains the following mass matrix for the doubly charged leptons
\begin{align}
	\mathscr{M}^{--}_{L_{\frac{3}{2}}\oplus E^a} = \bordermatrix{ & L_{\frac{3}{2}R}^{--} & E^{--}_R \cr 
		\overline{L_{\frac{3}{2}L}^{--}} & M_L & \sqrt2\,\lambda v \cr 
		\overline{E_L^{--}} & \sqrt2\lambar v &  M_E }.
\end{align}
Each mass matrix can be diagonalised  by introducing two appropriate
unitary transformations $U_L$ and $U_R$, in the form
\begin{subequations}\label{eq:SVD}
	\begin{align}
		U_L^{\nu\dagger} \mathscr{M}^0 U_R^\nu &= m_\nu \label{eq:SVD-neutral},\\ 
		U_L^{e\dagger} \mathscr{M}^- U_R^e &= m_e \label{eq:SVD-charged},\\
		U_L^{\rho\dagger} \mathscr{M}^{--} U_R^{\rho} &= m_\rho, \label{eq:SVD-doubleCharged}
	\end{align}
\end{subequations}
where $m_\nu, m_e$ and $m_\rho$ denote the diagonal mass matrices of the neutral, singly and doubly charged leptons.
In the following we will abbreviate $m_a\equiv m_{e_a}$ for the
charged lepton mass eigenvalues.
At the same time, mass-eigenstate fields are obtained by applying the
unitary transformations $U_L$ and $U_R$ onto the left- and
right-handed multiplets appearing in Eq.~(\ref{eq:mass-matrices}). We
denote the resulting mass-eigenstate fields as $\hat{\nu}_a,
\hat{e}_a, \hat{\rho}_a, \hat{\chi}_a$. Each of these carries lepton
number $+1$ and electric charge $Q=0,-1,-2,+1$, respectively.        
The number of fields in the multiplets for each seesaw
model is listed in Tab.~\ref{tab:field-content}.
\begin{table}[t]
	\centering
	\def\arraystretch{1.2}
	\begin{tabular}{c||c|c|c|c|c|c}
		\textbf{Model} & $L\oplus E$ & $L\oplus N$ & $L_{\frac{3}{2}}\oplus E$ & 
		$L\oplus E^a$ & $L\oplus N^a$ & $L_{\frac{3}{2}}\oplus E^a$ \\ \hline\hline
		$n_\nu$ ($Q=0$)\ & 4 & 5 & 3 & 5 & 5 & 4 \\
		$n_e$  ($Q=-1$)  & 5 & 4 & 5 & 5 & 5 & 5 \\
		$n_\rho$ ($Q=-2$)  & 0 & 0 & 1 & 1 & 0 & 2 \\
		$n_\chi$ ($Q=+1$) & 0 & 0 & 0 & 0 & 1 & 0 \\
	\end{tabular}
	\caption{number of mass eigenstates for the different electric
          charges $Q$ in the six VLL models.}
	\label{tab:field-content}
\end{table}
After the transformation the Yukawa interaction Lagrangian becomes
\begin{align}\label{eq:L-MB-Yukawa}
	\L_\text{Yuk} = - \frac{1}{\sqrt{2}}\sum_{f,\;\mathcal{S}=h,i\phi^0} \overline{f_a} Y^{\mathcal{S}}_{f,ab} \PR f_b \, \mathcal{S}
	\;-\; 
	\sum_{{ff' \;\mathcal{S}=\phi^+,\phi^-}} \overline{f_a} 
	Y^{\mathcal{S}}_{ff',ab} \PR f'_b \, \mathcal{S} ~+~ h.c.,
\end{align}
where $f,f'\in\{\hat{\nu}, \hat{e},\hat{\rho}, \hat{\chi}\}$ with $Q_f = Q_{f'}\pm1$ for $\mathcal{S}=\phi^\pm$. 
Similarly, the Lagrangian for the $Z$- and $W$-boson interactions is given by
\begin{align}\label{eq:L-MB-gauge}
	\L_\text{ZW} = \sum_{f} \overline{f_a} \slashed{Z} \Big[g^{Zf}_L \PL + g^{Zf}_R \PR\Big]_{ab} f'_b
	\;+\; 
	\bigg(\sum_{ff'} \bar{f_a}\slashed{W}^+ \Big[g^{Wff'}_L\PL + g^{Wff'}_R\PR\Big]_{ab} f'_b ~+~ h.c.\bigg).
\end{align}
with $Q_f = Q_{f'}+1$. Full expressions for the coupling matrices for
the different models are listed in App.~\ref{App:couplings}.\\

Finally, we briefly comment on the numerical computation of the mass-basis coupling matrices. The main complication arises from the
fact that, for a given set of BSM input parameters, the fundamental Yukawa couplings $y_i$ must be chosen to produce the correct SM lepton masses.
To obtain these values, we numerically solve the system of equations $\det{\mathcal{M}^{-\dagger}\mathcal{M}^--m_i^2}=0$ generated by the characteristic polynomials.
However, due to the possible complex phases this relation is ambiguous and requires an additional constraint. In our case, we use the perturbative expression
Eq.~\eqref{eq:lep-mass-perturbative} (setting $\theta_i=0$), discussed in more detail in the following section, as an initial guess for the solver which then further adjusts $\Re y_i$ to obtain a numerically exact root of the polynomials.

\subsection{Effective Couplings} \label{sec:eff-cpl}

In the following we give approximate results for the relevant effective couplings from Eqs.~\eqref{eq:L-MB-Yukawa}
and \eqref{eq:L-MB-gauge} corresponding to an expansion in the mass
suppression factors $v/M_{L,E}$. It is convenient to introduce the
dimensionless combinations\footnote{
	For the models $L\oplus N$ and $L\oplus N^a$ we also identify $\xi^E_i\equiv \lambda^N_i v/ M_N$ and $M_E\equiv M_N$ to avoid further notational clutter.}
\begin{align}\label{eq:cpl-abbr}
	\xi^L_i \equiv \frac{\lambda^L_i v}{M_L} , \qquad
	\xi^E_i \equiv \frac{\lambda^E_i v}{M_E},
\end{align}
of the mass suppression factors and  the BSM Yukawa couplings
that will naturally appear in the following expressions.
\begin{figure}[t]
	\centering
	\begin{subfigure}{.24\textwidth}
		\centering
		\includegraphics[width=\textwidth]{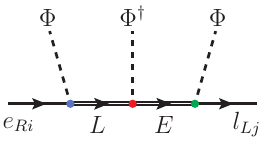} 
		\caption{}
		\label{fig:HHH}
	\end{subfigure}\hfill
	\begin{subfigure}{.49\textwidth}
		\includegraphics[width=.49\textwidth]{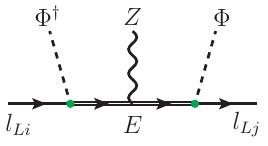}
		\includegraphics[width=.49\textwidth]{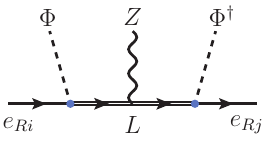}
		\caption{}
		\label{fig:HVH}
	\end{subfigure}\hfill
	\begin{subfigure}{.24\textwidth}
		\includegraphics[width=\textwidth]{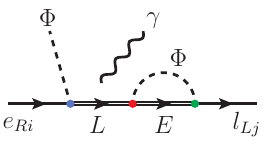}
		\caption{}
		\label{fig:HV}
	\end{subfigure}
	\caption{Example of diagrams in the $L\oplus E$ model inducing the mass 
		suppressed corrections to the SM couplings after EWSB. 
		\textbf{(a)} tree-level correction to Higgs coupling and lepton mass.
		\textbf{(b)} tree-level correction to left- and right-handed lepton--$Z$ coupling.
		\textbf{(c)} one-loop (leading order) chirally enhanced correction to the lepton form factors.
	}
	\label{fig:SMEFT-diagrams}
\end{figure}

\begin{table}[t]
	\centering
	\def\arraystretch{1.2}
	\begin{tabular}{c||c|c|c|c|c|c}
		\textbf{Model} & $L\oplus E$ & $L\oplus N$  & $L_{\frac{3}{2}}\oplus E$  & $L\oplus E^a$ & $L\oplus N^a$  & $L_{\frac{3}{2}}\oplus E^a$ \\ \hline\hline
		$k_{LR}$ & $1$   & $0$  & $1$  & $1$  & $2$  & $-1$ \\
		$k_L$    & $-\frac{1}{2}$  & $0$ & $-\frac{1}{2}$ & $-\frac{1}{2}$ & $-1$ & $-\frac{1}{2}$ \\
		$k_R$    & $-\frac{1}{2}$  & $-\frac{1}{2}$  & $-\frac{1}{2}$ & $-\frac{1}{2}$ & $-\frac{1}{2}$ & $-\frac{1}{2}$ \\ \hline
		$q_L$ & $1$ & $0$ & $1$ & $1$ & $-2$ & $1$ \\
		$q_R$ & $-1$ & $-1$ & $1$ & $-1$ & $-1$ & $1$ \\ 
		$w_L$ & $-1$ & $-1$ & $-1$ & $1$ & $1$ & $1$ \\	\hline
		$c_{LR}$ & $-1$        & $0$         & $-5$       & $-9$        & $-2$        & $+5$ \\
		$c_L$    & $8s_W^2-1$  & $-6$        & $8s_W^2-1$ & $8s_W^2+27$ & $-16s_W^2$  & $8s_W^2+27$ \\
		$c_R$    & $-8s_W^2+11$ & $-8s_W^2+11$ & $8s_W^2-1$ & $-8s_W^2+11$ & $-8s_W^2+11$ & $8s_W^2-1$
\\ \hline
$\mathcal{Q}$ & 1 & -- & 5 & 9 & 1 & 5
	\end{tabular}
	\caption{Coefficients of the leading-order corrections to the
          effective couplings in the different VLL models. For later
          convenience we also list the values of $\mathcal{Q}= -{c_{LR}}/{k_{LR}}$.}
	\label{tab:cpl-coeff}
\end{table}

We begin with the lepton mass eigenvalues.
The off-diagonal terms in the lepton mass matrices change the fundamental relation between the SM lepton masses $m_i$ and
Yukawa couplings $y_i$. The singular value decomposition Eq.~\eqref{eq:SVD} can be solved perturbatively in powers of $v$, yielding
the relation
\begin{align}\label{eq:lep-mass-perturbative}
	m_i e^{i\theta_i} = y_i v \Big(1 + k_L |\xi^E_i|^2 + k_R |\xi^L_i|^2\Big) + k_{LR} \lambar\xi^E_i\xi^L_i v + \O(\xi^4)
\end{align}
where $m_i$ is the resulting tree-level mass,  
$\theta_i$ denotes the overall phase of the right-hand side, and the
coefficients $k_X$ for the different models are listed in
Tab.~\ref{tab:cpl-coeff}. The phase $\theta_i$ could be absorbed by a
redefinition of the mass-eigenstate fields in the operator $\overline{\hat{e}_{Li}}\hat{e}_{Ri}$
and thus affects
all other chirality-flipping observables.

The terms in Eq.~(\ref{eq:lep-mass-perturbative}) can be understood via mass-insertion
Feynman diagrams. The diagram in
Fig.~\ref{fig:HHH} effectively generates a dimension-6 operator of the form\footnote{The actual SU$(2)_L$ contractions
	of the operators differ for the different VLL representations.} $\overline{l_L}e_R\Phi(\Phi^\dagger\Phi)$
that (after EWSB) yields the term $\propto \lambar$. Diagrams such as
Fig.~\ref{fig:HVH} induce operators of the form  
$(\overline{l_L}\Phi)\slashed{D}(\Phi^\dagger l_L)$ and $\overline{e_R}\slashed{D}e_R(\Phi^\dagger\Phi)$ that contribute to 
the field normalization and subsequently produce the
$|\xi|^2$-corrections to the SM-like mass term.

Crucially, the contributions to the lepton--Higgs couplings differ
from the contributions to the masses by combinatorial factors. Hence,
after eliminating the SM-like Yukawa couplings $y_i$ in favour of the
lepton masses $m_i$ using Eq.~(\ref{eq:lep-mass-perturbative}), there
remain non-trivial corrections to the lepton--Higgs coupling given 
by
\begin{align}\label{eq:Higgs-coupling}
	Y^h_{e,ij} = \tfrac{m_i}{v} \delta_{ij} + 
	2 \Big(k_{LR} \,\lambar\, \xi^E_i\xi^L_j + k_L \xi^E_i\xi^{E*}_j\frac{m_j}{v}  + k_R \frac{m_i}{v} \xi^{L*}_i\xi^L_j + \O(\xi^4) \Big)  e^{-i\theta_i}.
\end{align}
The appearing terms can again be understood via diagrams such as the ones in
Fig.~\ref{fig:HHH} and \ref{fig:HVH}.

Similarly, the lepton--$Z$ interactions can be approximated as
\begin{subequations}\label{eq:Z-coupling}
	\begin{alignat}{3}
		g^{Ze}_{L,ij} &= ~ &\frac{g_W}{c_W} \bigg( &\Big[s_W^2-\frac{1}{2}\Big] \delta_{ij}& &+  \frac{q_L}{2} \xi^E_i\xi^{E*}_j + \O(\xi^4)\bigg), \\
		g^{Ze}_{R,ij} &= ~ &\frac{g_W}{c_W} \bigg(  &\quad s_W^2 \delta_{ij}&  &+  \frac{q_R}{2} \xi^{L*}_i\xi^L_j
		+\O(\xi^4)\bigg),
	\end{alignat}
\end{subequations}
where again diagrams such as the ones in Fig.~\ref{fig:HVH} are the
origin of the $\xi\xi^*$-type corrections.

The analogous expressions for the $W$-boson couplings are given by
\begin{align}\label{eq:W-coupling}
	g^{W\nu e}_{L,ii} = \frac{g_W}{\sqrt2} \bigg( 1 + \frac{w_L}{2} |\xi^E_i|^2 + \O(\xi^4) \bigg), \qquad g^{W\nu e}_{R,ij}=0,
\end{align}
where the singlets $E$ and $N$ yield $w_L=-1$ and the triplets $E^a$ and $N^a$ give $w_L=+1$.
The off-diagonal entries have a more complicated structure compared to Eq.~\eqref{eq:Z-coupling}, due to the unrelated
neutrino and electron diagonalization matrices. While they will be irrelevant in the following discussion, 
their full calculation can be derived from the coupling matrices listed in App.~\ref{App:couplings}.

Finally, the lepton photon interaction is modified at one-loop order.
We write the quantum-corrected interaction vertex of leptons with the
photon using the following covariant decomposition (see e.g.~Refs.~\cite{Jegerlehner:2009ry,Athron:2025ets})
\begin{align}\label{eq:cov-dec-photon-vertex}
	\begin{gathered}
		\includegraphics[scale=.9]{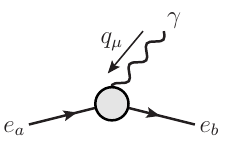}
	\end{gathered} = -i Q_e e \bar u_b(p') \left[\gamma^\mu \delta_{ab} F^a_E  + \frac{i\sigma^{\mu\nu}q_\nu}{m_a+m_b} (F_M^{ba} - i F_D^{ba} \gamma^5)\right] u_a(p) \epsilon^*_\mu(-q) ,
\end{align}
where the on-shell conditions of the external leptons enforce
$p^2=m_a^2$ and $p'^2=m_b^2$, while the form factors are real and
depend only on  $q^2$, $F_i=F_i(q^2)$. In the application to dipole
moments as well as $e_i\to e_j\gamma$, the momentum $q=0$. In cases
where the interaction vertex is just a subdiagram (like e.g.\ in
$\mu\to 3e$), $q^2$ can in principle be non-zero. However, in our applications the momentum dependence can be neglected.
The full one-loop results using mass eigenstates are given in App.~\ref{App:FF}; here we show the leading result in the limit of large VLL masses
\begin{align}\label{eq:form-factors}
	F_M^{ij}(0) - i F_D^{ij}(0) = \frac{m_i+m_j}{32\pi^2 v} \bigg(c_{LR} \, \xi^E_i\lambar\xi^L_j + 
	\frac{c_L}{6}\xi^E_i\xi^{E*}_j\frac{m_j}{v} + \frac{c_R}{6} \frac{m_i}{v}\xi^{L*}_i\xi^L_j  + \O(\xi^4) \bigg) e^{-i\theta_i}
\end{align}
where the coefficients are listed in Tab.~\ref{tab:cpl-coeff}. The
appearing terms are very similar to the ones in the previous
expressions and can be understood similarly. Particularly the
chirally enhanced $\lambar$ term corresponds to the diagram shown in Fig.~\ref{fig:HV}.

\section{Flavour independent constraints}\label{sec:Constraints}

Here we consider a set of basic constraints on the VLL parameters. For
our later purposes, limits on the Yukawa couplings from electroweak
precision observables (EWPO) are most relevant, but we also briefly
discuss constraints from LHC searches and from vacuum stability.

\paragraph{Electroweak precision observables}

\begin{figure}[t]
	\centering
	\begin{subfigure}{.28\textwidth}
		\includegraphics[width=\textwidth]{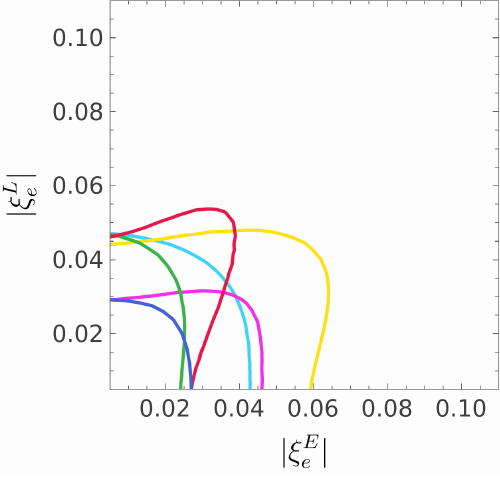}
		\caption{electron exclusion contours}\label{fig:EWPO-e}
	\end{subfigure}\hfill
	\begin{subfigure}{.28\textwidth}
		\includegraphics[width=\textwidth]{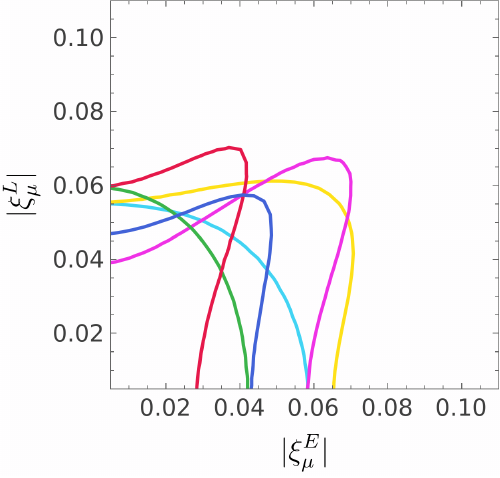}
		\caption{muon exclusion contours}\label{fig:EWPO-mu}
	\end{subfigure}\hfill
	\begin{subfigure}{.28\textwidth}
		\includegraphics[width=\textwidth]{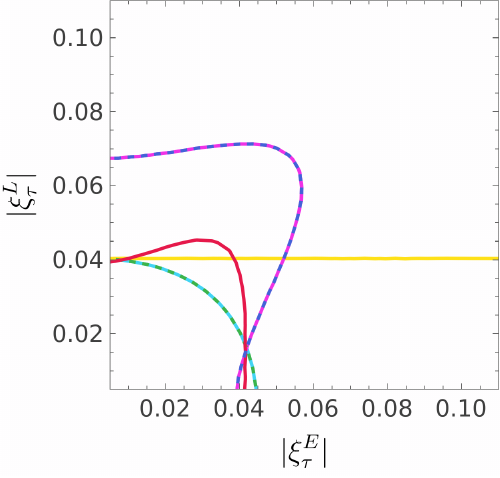}
		\caption{tau exclusion contours}\label{fig:EWPO-tau}
	\end{subfigure}\hfill
	\begin{subfigure}{.1\textwidth}
		\centering
		\raisebox{.6\height}{\includegraphics[width=\textwidth]{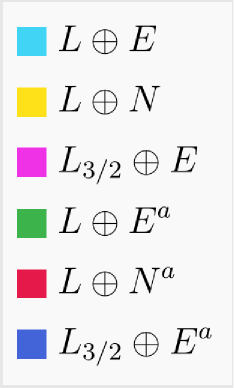}}
	\end{subfigure}
	\caption{2$\sigma$ EWPO exclusion limits on VLL couplings to the \text{(a)} electron, \textbf{(b)} muon
	and \textbf{(c)} tau, where the couplings not shown are set to 0 respectively.}
	\label{fig:EWPO-excl}
\end{figure}

For the evaluation of EWPOs we use a scheme where the muon decay
  constant $G_\text{F}$, the $Z$-boson mass $M_Z$ and the Thomson
  limit finestructure constant are used as input, while $M_W$ and
  effective $Z$-couplings are treated as predictions, as described
  e.g.~in Ref.~\cite{ALEPH:2005ab} and references therein. In this
  context, the tree-level contribution to the lepton--$W$ couplings
  Eq.~\eqref{eq:W-coupling} implies a tree-level BSM correction to the
  quantity $\Delta r$ appearing in the 
  relationship between the input parameters and $M_W$,
\begin{align}
  \delta(\Delta r)\approx
  \frac{w_L}{2} \Big(|\xi^E_1|^2 + |\xi^E_2|^2\Big).
\end{align}
This in turn
implies  shifts of $M_W^2$, $s_W^2$ and $g_W$ given by
$\delta_W \equiv \frac{\delta M_W^2}{M_W^2} =
\frac{s_W^2}{s_W^2-c_W^2} \delta(\Delta r)$,
$\delta c_W^2 = c_W^2 \delta_W = - \delta s_W^2$ and $\delta g_W=g_W
\delta c_W^2/(2s_W^2)$, compared to the respective SM predictions.

On the one hand,
$\delta_W$ is constrained by the current experimental value $M_W^\text{exp} = 80.3692\pm0.0133$ GeV \cite{ParticleDataGroup:2024cfk},
compared to the SM prediction $M_W^\text{SM}=80.356\pm 0.006$ GeV \cite{deBlas:2021wap,ParticleDataGroup:2024cfk}\footnote{
The $M_W^{\text{exp}}$ quoted in the main text does not include the recent CDF $M_W$ measurement \cite{CDF:2022hxs} in the averaging procedure. That CDF measurement yields a significantly higher value for $M_W$ that is in tension with other measurements. In the main text we will not use the CDF measurement, but we remark that the singlet models yield a positive contribution to $M_W$
that could match the CDF value for $|\xi^E_i|\sim 0.1$ which, however, is in strong tension with the $Z$-pole constraints.
On the other hand, the triplet models reduce $M_W$ compared to the SM prediction and therefore increase the tension with the CDF result.
 }. 

On the other hand, these shifts affect the predictions for the
pseudo-observables $g_{L/R,ii}$ given in Tab.~7.7 and Tab.~G.3 of
Ref.~\cite{ALEPH:2005ab},\footnote{%
Similar constraints from NuTeV data on electroweak neutrino
interactions have been considered in Ref.~\cite{Freitas:2025bgg}, and
we thank A.~Freitas for pointing out that reference and for discussions.
}
which correspond to the effective lepton-$Z$
couplings of  Eq.~\eqref{eq:Z-coupling}, where the prefactor $g_W/c_W$
is split off in $G_\text{F}M_Z^2$ parametrisation. At tree-level, these quantities are
given as
\begin{align}
  g_{L/R,ii} &= \bigg(\frac{c_W}{g_W}\bigg) \left(1-\frac{\delta(\Delta r)}{2}\right) g^{Ze}_{L/R,ii}.
\end{align}
Compared to the full SM prediction, there are thus additional tree-level
BSM contributions
\begin{align}
  \delta   g_{L,ii} &\approx  -\frac{\delta_W}{4s_W^2} +  \frac{q_L}{2}
  \xi^E_i\xi^{E*}_j ,
  &
  \delta g_{R,ii} &  \approx - \frac{\delta_W}{2} +  \frac{q_R}{2} \xi^{L*}_i\xi^L_j.
\end{align}
With these ingredients we can construct the differences $\Delta M_W = M_W^\text{SM} (1+\frac{1}{2}\delta_W) - M_W^\text{exp}$
and $\Delta g_i = g_{L/R,ii}^\text{SM}+\delta g_{L/R,ii} - g_{L/R,ii}^{\text{exp}}$
together with the corresponding $\chi^2$ test
\begin{align}
	\chi^2 = \frac{\Delta M_W^2}{\sigma_W^2} +  \Delta g^T \Sigma^{-1} \Delta g,
\end{align}
where $\sigma_W\approx 0.015$ GeV and the covariance matrix is given by $\Sigma = \Sigma^\text{exp} + \Sigma^\text{theory}$ where
$\Sigma_{ij}^\text{exp}=\sigma_i\rho_{ij}\sigma_j$ is obtained from
the correlation matrix $\rho$ and experimental uncertainties $\sigma$
given in Ref.~\cite{ALEPH:2005ab} (Tab. 7.7). The theory covariance 
is given by $\Sigma^\text{theory}_{ij} = \sigma_i\sigma_j$, where $\sigma_i$
is obtained from the SM uncertainties (parametric, fully correlated) listed in Tab.~G.3 of Ref.~\cite{ALEPH:2005ab} 
while uncertainties from missing higher-order corrections are neglected. 

To obtain bounds on the couplings, we then consider $\Delta\chi^2\equiv\chi^2 - \chi^2_\text{min}$ where the best-fit value of the
(possibly constrained) parameter-space is subtracted. This value is approximately $\chi^2_{6-c}$ distributed where $c$ denotes the 
number of additional constraints (like e.g. $\xi^E_i=0$).

Fig.~\ref{fig:EWPO-excl} shows 2$\sigma$ exclusion contours from $\Delta\chi^2$ in the $\xi^E$--$\xi^L$ plane for VLL coupling exclusively to the electron, muon and
tau, respectively. 
The resulting bounds agree qualitatively with those obtained in previous studies for muon-specific couplings \cite{Kannike:2011ng,Crivellin:2020ebi}.
In case of simultaneous couplings to multiple generations, the contours will change slightly, however, in this case constraints from LFV observables
are typically much stronger. This point will be discussed in detail in Sec.~\ref{sec:LFV}. \\

Loop contributions to EWPOs are frequently considered in studies of VLL \cite{Zhang:2006de,Cynolter:2008ea,Joglekar:2012vc,Ellis:2014dza,Freitas:2014pua,Cingiloglu:2024vdh} via the electroweak
oblique parameters $S$, $T$ and $U$. These are  induced at one-loop
order via VLL corrections to the gauge-boson self-energies
\begin{subequations}
	\begin{align}
		\alpha S &= \frac{4s_W^2c_W^2}{M_Z^2} \bigg(\Pi_{ZZ}(M_Z^2) - \Pi_{ZZ}(0) - \frac{c_W^2-s_W^2}{c_W s_W}\Pi_{ZA}(M_Z^2) - \Pi_{AA}(M_Z^2)\bigg), \\
		\alpha T &= \frac{\Pi_{WW}(0)}{M_W^2} - \frac{\Pi_{ZZ}(0)}{M_Z^2}, \\
		\alpha U &= \frac{4s_W^2}{M_W^2} \bigg( \Pi_{WW}(M_W^2) - \Pi_{WW}(0) - \frac{c_W^3}{s_W} \Pi_{ZA}(M_Z^2) - c_W^2 \Pi_{AA}(M_Z)^2 \bigg),
	\end{align}
\end{subequations}
with the current best fit values for the difference between the NP and SM values given by
$\Delta S=0.021\pm0.096$, $\Delta T=0.04\pm 0.12$ and $\Delta U=0.008\pm 0.092$ \cite{ParticleDataGroup:2024cfk}.
While these constraints were relevant in case of light $\O(100~\text{GeV})$ VLL, they become 
significantly weaker than the lepton-$Z$ bounds in case of $\O(1~\text{TeV})$ masses relevant to the discussion in this paper.
Furthermore, while they illustrate the general behaviour, the oblique parameters are insufficient to fully capture all relevant loop corrections.

It is nonetheless interesting to point out the different parametric
scaling of these quantities. For example, in the $L\oplus E$ model (and with $\lambda=\lambar=0$) the $T$ parameter is given by
\begin{align}
	\Delta T = \frac{M_E^2}{32\pi^2 v^2 \alpha} \Big(\sum_i |\xi^E_i|^2\Big)^2 + 
	\frac{M_L^2}{24\pi^2 v^2\alpha} \Big(\sum_i |\xi^L_i|^2\Big)^2 + \O(\xi^6).
\end{align}
Despite the additional enhancement from the Dirac masses,
due to the loop suppression and the $|\xi|^4$ scaling 
the bound becomes stronger than the one from lepton--$Z$ couplings only for $M_{L/E}\gtrsim \O(10\,\text{TeV})$.
On the other hand, unlike the lepton--$Z$ couplings, the oblique
parameters are also affected by $\lambda$ and $\lambar$ at leading
order and thus, in principle, lead to constraints on these parameters.
However, for our purposes, the bounds obtained from $S$, $T$ and $U$ are weaker
than other considered constraints and hence not discussed further.

Beyond providing bounds on the parameter space, future improvements in the experimental sensitivities of 
both the lepton--$Z$ couplings and oblique parameters may also offer a criterion for discriminating between the models.
This point will be further discussed in Sec.~\ref{sec:patterns}.

\paragraph{LHC searches}
Several searches for VLLs and heavy neutral or multi-charged particles have been performed by
the ATLAS \cite{ATLAS:2023sbu,ATLAS:2024mrr,ATLAS:2025cdi,ATLAS:2025wgc} and CMS \cite{CMS:2019hsm,CMS:2022nty,CMS:2022cpe,CMS:2024bni,CMS:2024xdq} collaborations.
While these searches have so far focused exclusively on the $E$ and $L$ representations, the 
considered production $pp\to W^* \to L^- N^0$, $pp\to Z^*/\gamma^* \to L^+L^-$ and decay channels
$L\to\nu W, eZ,eh$ and $N^0\to eW$ exists also for the other representations, 
such that the exclusion limits approximately apply in all of the scenarios.
All of these searches point towards VLL masses below
$\O(1~\text{TeV})$ being excluded. For the purposes of the present paper
the exact lower bounds are not very important as the phenomenology depends on the masses mainly via $\xi_i$ where any increase in
the VLL masses can be compensated by a corresponding increase in the Yukawa couplings $\lambda_i$. Nonetheless, for the following
analyses we will choose VLL masses that are generally safely above the current exclusion limits.

\paragraph{Vacuum Stability}\label{sec:vacuum}
Finally we comment on considerations of vacuum stability. When VLL
fields couple to the Higgs via Yukawa couplings, the loop-corrected
Higgs potential is modified; for large Yukawa couplings the effective
running quartic Higgs coupling can become negative already at rather
moderate energy scales. Refs.~\cite{Xiao:2014kba,Blum:2015rpa,Gopalakrishna:2018uxn,Hiller:2022rla,Arsenault:2022xty,Cingiloglu:2024vdh,Adhikary:2024esf} have investigated
vacuum stability in VLL models and found that generally $|\lambda_i|
\gtrsim 1$ tend to be excluded. While a more elaborate study including
higher-order corrections would be motivated in view of intriguing
phenomenology that is possible at large $\lambda_i$ (see
Refs.~\cite{Mohling:2024qvk} and later in
Sec.~\ref{sec:LFC}), for most of the discussions in the
present paper it is sufficient to focus on the parameter region with
$|\lambda_i|\lesssim 0.5$ that is not endangered by vacuum
stability.

Notably, in contrast to the electroweak precision constraints, the vacuum stability places bounds directly on the 
Yukawa couplings as opposed to bounds on the $\xi$'s. Compared to the
bounds in Fig.~\ref{fig:EWPO-excl}, the vacuum stability requirement
thus becomes stronger at VLL masses $\gtrsim 2$ TeV.

\section{Phenomenology of flavour conserving observables}\label{sec:LFC}

The defining feature of the six considered VLL models are the
seesaw-like lepton mass terms, in addition to the usual SM-like 
Yukawa term. These lead to peculiar
and correlated contributions to many chirality-flipping observables,
as illustrated already in Eq.~(\ref{previewVLLrelations}).
Here we expand the analysis of chirality-flipping and
flavour-conserving observables, i.e.\ of the lepton masses $m_i$, the
lepton--Higgs couplings $\lambda_{ii}$ and the magnetic and electric
dipole moments $a_i$ and $d_i$. 

The impact of VLL on such flavour-conserving observables has already been extensively studied in the past,
particularly in context of $\Delta a_\mu$ and the Higgs measurements at the LHC 
\cite{delAguila:2008pw,Kannike:2011ng,Ishiwata:2011hr,Dermisek:2013gta}, motivated by the strong correlation
already highlighted in Eq.~\eqref{previewVLLrelations}. Since these analyses significant progress
in both the experimental measurements and SM predictions has been made. We therefore expand these
discussions here and update the conclusions for the relevant parameter regions of the VLL models. 

Most notably, the reevaluation of the hadronic vacuum polarisation
contribution to $a_\mu^\text{SM}$ based on new lattice QCD
calculations  
has shifted the central value of the SM prediction as determined in
the Theory Initiative White Paper Ref.~\cite{Aliberti:2025beg} based on original Refs.~\cite{\SMref}. At the same time, the Fermilab $g-2$ experiment
has published its final result, surpassing the expected precision goal while staying in full agreement with previous
measurements \cite{\expref}. The combined new central value of the deviation is given by
\begin{align}
	\Delta a_\mu^{\text{Exp$-$WP2025}} &= (3.8\pm 6.3) \times 10^{-10}, \label{eq:WP25} \\
	\intertext{which is fully compatible with zero and differs
          significantly from the previous value based on the first
          White Paper Ref.~\cite{Aoyama:2020ynm},}
	\Delta a_\mu^{\text{Exp$-$WP2020}} &= (26.2\pm4.5) \times 10^{-10}. \label{eq:WP20}
\end{align}
On the other hand, the $h\to\mu\mu$ signal strength modifier has been determined by both the 
ATLAS \cite{ATLAS:2020fzp} and CMS \cite{CMS:2020xwi,CMS:2026nce} collaborations yielding the current PDG average \cite{ParticleDataGroup:2024cfk}
\begin{align}\label{eq:muon-signal-strength}
	\mu_{\mu\mu} = \frac{\sigma\cdot\Gamma(h\to\mu\mu)}{[\sigma\cdot\Gamma(h\to\mu\mu)]_\text{SM}} = 1.21\pm 0.35,
\end{align}
also in good agreement with the SM. Other current measurements of flavour-conserving observables considered in this paper
together with sensitivity estimates of future experiments are listed in Tab.~\ref{tab:LFCbounds}.
The theory predictions for the
dipole moments are given as 
$\Delta a_i = F_M^{ii}(0) - F_M^{ii}(0)_\text{SM}$ and
$d_i= \tfrac{e}{2m_i} F_D^{ii}(0)$.
Since the VLL do not significantly change the Higgs production cross
section, the signal strength modifier is given as
 $\mu_{ii}\approx R_{ii}\equiv \frac{\Gamma(h\to e_i \bar{e}_i)}{\Gamma(h\to e_i \bar{e}_i)_\text{SM}}
	= \Big|\tfrac{\lambda_{ii}}{\lambda_{ii}^\text{SM}}\Big|^2 $, and the effective lepton--Higgs
coupling $\lambda_{ii}$ at tree-level can be identified with the
coupling factor $Y^h_{e,ii}$ in Eq.~(\ref{eq:Higgs-coupling}), while its
SM counterpart is $\lambda_{ii}^{\text{SM}}=m_{i}/v$.
While full analytical results are discussed in Sec.~\ref{sec:Models}
and the Appendix, we provide here compact approximations,
highlighting the essential behaviour governed by chirally enhanced
terms $\propto\lambar$. In this approximation, the two contributions
to the lepton masses in Eq.~(\ref{eq:lep-mass-perturbative}) can be
written as \cite{Kannike:2011ng,Dermisek:2013gta,Dermisek:2023nhe}
\begin{subequations}
	\begin{align}
	 m_i^{H} &=y_i v,
	 \\
	 m_i^{HHH} &= k_{LR} \lambar\xi^E_i\xi^L_i v,
	\end{align}
\end{subequations}
where $k_{LR}$ is a half-integer that depends on the model, see
Tab.~\ref{tab:cpl-coeff} . 
Eq.~(\ref{eq:lep-mass-perturbative})
contains additional contributions that are irrelevant for the purposes
here but that will become more relevant in the context of
lepton-flavour violating observables discussed in the following section.

\begin{figure}[t]
	\centering
	\includegraphics[width=.3\textwidth]{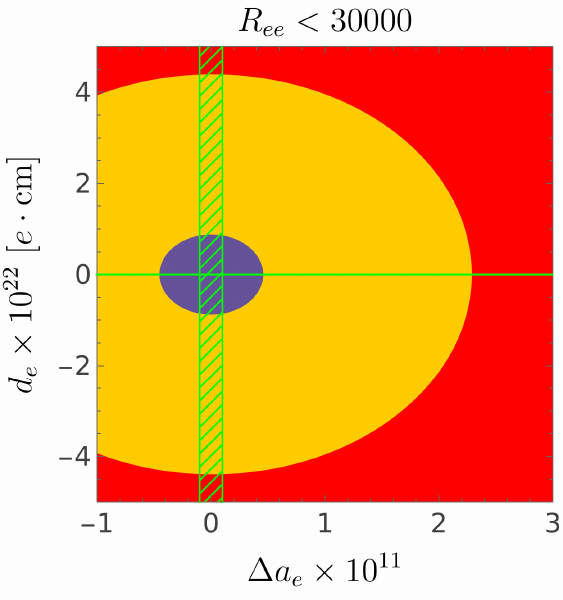} \hfill
	\includegraphics[width=.3\textwidth]{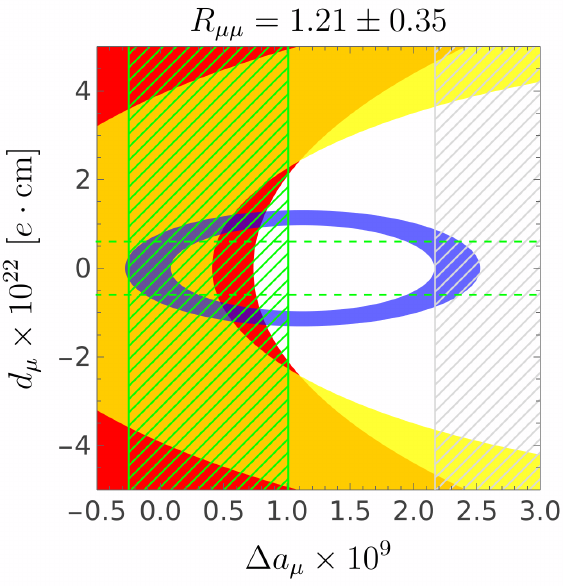} \hfill
	\includegraphics[width=.3\textwidth]{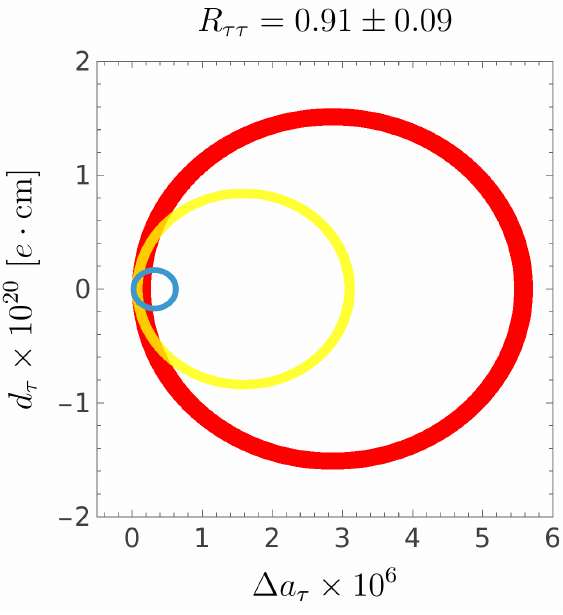}
	\caption{Current constraint on the lepton dipole moments from Eq.~\eqref{eq:Rii-correlation} for the different values
		of $\mathcal{Q}=1$ (blue), $5$ (yellow) and $9$ (red). The green hatched regions show the current constraint
		from dipole measurements, the dashed lines show the future sensitivity of the muon EDM experiment at PSI.
		The gray hatched region shows the old bounds from Eq.\eqref{eq:WP20}. }
	\label{fig:dipoleellipses}
\end{figure}

Importantly, the contribution $m^{HHH}_i$ represents an additional source of chiral symmetry breaking in the lepton sector
and contributes to all chirality-flipping observables. 
With these definitions, the correlations previewed already in the
Introduction in Eq.~(\ref{previewVLLrelations}) are explained via
Eqs.~(\ref{eq:lep-mass-perturbative},\ref{eq:Higgs-coupling},\ref{eq:form-factors}).
Combining these relations leads to the correlation \cite{Dermisek:2022aec} 
	\begin{align}\label{eq:Rii-correlation}
		R_{ii} = \bigg(1 - \frac{\Delta a_i}{2\omega_i\mathcal{Q}}\bigg)^2 + 
		\bigg(\frac{m_i d_i}{e\omega_i \mathcal{Q}}\bigg)^2,
	\end{align}
with $\omega_i = m_i^2 / (64\pi^2 v^2)$ and $\mathcal{Q}= -{c_{LR}}/{k_{LR}}$. For fixed $R_{ii}$ this equation describes an ellipse in
the $\Delta a_i$ -- $d_i$ plane. While a correlation of this form can be found in a variety of models \cite{Dermisek:2022aec,Dermisek:2023nhe}
(see also \cite{Crivellin:2021rbq,Athron:2025ets}), the notable distinction of the VLL models is the relative loop suppression of the dipole moments
compared to the Higgs coupling corrections, stemming from the fact that the latter is generated already at tree-level.
As a consequence, the normalization factor $\omega_i$ has the typical order of magnitude for electroweak corrections to the dipole moments,
leading to a very strong correlation. Fig.~\ref{fig:dipoleellipses} shows the resulting constraint from the current values of $R_{ii}$
for all three lepton generations. From Tab.~\ref{tab:cpl-coeff} one finds only three distinct values of 
$\mathcal{Q}=1$ ($L\oplus E$ and $L\oplus N^a$), $\mathcal{Q}=5$ ($L_{3/2}\oplus E$ and $L_{3/2}\oplus E^a$) and
$\mathcal{Q}=9$ ($L\oplus E^a$). One special case is $L\oplus N$, where the chirally enhanced contributions \emph{accidentally} vanish, due to the antisymmetric SU$(2)$ contraction in the operator $\overline{l_L}e_R\Phi(\tilde\Phi^\dagger\Phi)$ induced by the analogous diagram to Fig.~\ref{fig:HHH}.

The three plots illustrate the special role played by the muon. Here,
all three constraints on $R_{\mu\mu}$, $a_\mu$ and $d_\mu$ imply
similarly strong restrictions of parameter space. In contrast, for the
electron, $R_{ee}$ is essentially unconstrained but the limits on
$a_e$ and $d_e$ are very strong. For the $\tau$-lepton, $R_{\tau\tau}$
is tightly constrained, leading to very specific viable combinations
of $a_\tau$ and $d_\tau$ in each VLL model. Reaching the necessary precision 
in order to test this correlation will require a significant experimental effort,
but may optimistically be possible with sufficient statistics \cite{Crivellin:2021spu,USBelleIIGroup:2022qro}.

Lastly, we specialise to the case without CP violation (corresponding to real couplings and $d_i=0$) and focus specifically on the muon.
In this case, the constraint Eq.~\eqref{eq:muon-signal-strength} together with the correlation Eq.~\eqref{eq:Rii-correlation} singles out 
two cases, either (i) $\Delta a_\mu \approx 0$ and
$\lambda_{\mu\mu}\approx \lambda_{\mu\mu}^\text{SM}$ (corresponding to
the left apex of the ellipses in the middle plot of Fig.~\ref{fig:dipoleellipses}), or
(ii) $\Delta a_\mu \approx 4\mathcal{Q}\omega_\mu$ resulting in
$\lambda_{\mu\mu}\approx -\lambda_{\mu\mu}^\text{SM}$  (corresponding to
the right apex of the ellipses).
Explicitly, the allowed ranges are (see also Fig.~5.20
in Ref.~\cite{Athron:2025ets})
\begin{align}
	\Delta a_\mu = \mathcal{Q}\times 10^{-10} \times \begin{cases}
		-1.1\pm1.8 & \lambda_{\mu\mu} \approx + \lambda_{\mu\mu}^\text{SM}, \\
		\,23.6\;\!\pm1.8 & \lambda_{\mu\mu} \approx - \lambda_{\mu\mu}^\text{SM}.
	\end{cases}
\end{align}
For $\mathcal{Q}=1$, the second scenario happens to agree quite well with the old deviation Eq.~\eqref{eq:WP20}, as discussed in
Refs.~\cite{Dermisek:2013gta,Athron:2025ets}. However, the updated result Eq.~\eqref{eq:WP25} no longer supports this scenario\footnote{%
	Higher-order corrections in this regime can however be significant
	\cite{Mohling:2024qvk} and potentially modify this conclusion. A
	discussion of these effects is, however, beyond the scope of the present paper.} in
any of the models and instead prefers $\Delta a_\mu\approx 0$. 

This in turn motivates to focus on the region where the additional
chiral symmetry breaking is moderate, corresponding to smaller
preferred values of the parameter $\lambar$. Indeed, by exhausting the
maximum $\xi$-values allowed by the
electroweak constraints in Fig.~\ref{fig:EWPO-excl}, the
maximum possible contributions to $a_\mu$ in the six models are
\begin{align}
	|\Delta a_\mu| \lesssim |c_{LR} \lambar|
        \times
	4.5\times 10^{-9}.
\end{align}
The previously motivated opposite-sign case and the result of Eq.~(\ref{eq:WP20})  thus required values such
as $\lambar\sim0.5$ (or more, if the $\xi$-values do not exhaust
Fig.~\ref{fig:EWPO-excl}). In contrast, the current preferred
value for $a_\mu$ in Eq.~(\ref{eq:WP25}) does not require such
values. We will therefore restrict our attention to parameter regions with
\begin{align}\label{lambarconstraint}
  |\lambar|<0.5
\end{align}
in the following phenomenological investigations. This restriction is further
motivated by the discussion of vacuum stability in Sec.~\ref{sec:vacuum}, and the
fact that with Eq.~(\ref{lambarconstraint}), $\lambar$ is in a similar
range as the other Yukawa couplings both in the SM lepton sector and
in the VLL sector.

\begin{table}[t]
	\centering
	\def\arraystretch{1.2}
	\begin{tabular}{c|cc|cc}
		\textbf{Observable} & \multicolumn{2}{c|}{\textbf{current bound}} & \multicolumn{2}{c}{\textbf{future sensitivities}} \\ \hline\hline
		$|\Delta a_e|$ & $< 10^{-12}$ & \footnotemark & -- &\\
		$\Delta a_\mu$ & $\SI{3.8(6.3)e-10}{}$ &\cite{Aliberti:2025beg} & -- &\\ \hline
		$|d_e|$ & $< \SI{4.1e-30}{\ecm}$ & \cite{Roussy:2022cmp} & $<\SI{1e-31}{\ecm}$ & \cite{Fitch:2020jil} \\
		$|d_\mu|$ & $< \SI{1.9e-19}{\ecm}$ & \cite{Muong-2:2008ebm} & $<\SI{6e-23}{\ecm}$ & \cite{Sakurai:2022tbk}\\ \hline
		$R_{ee}$ & $ < \SI{30000}{}$ & \cite{CMS:2022urr,ATLAS:2019old}& $< 2.6$  & \cite{deBlas:2019rxi,dEnterria:2021xij} \\
		$R_{\mu\mu}$ & $\SI{1.21(0.35)}{}$ & \cite{ParticleDataGroup:2024cfk} &
		$\delta R_{\mu\mu} \approx 0.8\%$ & \cite{FCC:2025lpp} \\
		$R_{\tau\tau}$ & $\SI{0.91(0.09)}{}$ & \cite{ParticleDataGroup:2024cfk}& 
		$\delta R_{\tau\tau} \approx 0.8\%$ & \cite{FCC:2025lpp}
	\end{tabular}
	\caption{List of current bounds and future sensitivities for flavour-conserving observables. }
	\label{tab:LFCbounds}
\end{table}	
\footnotetext{For
	$\Delta a_e$, a conservative approximate bound based on the
	measurements of $a_e$ \cite{Hanneke:2008tm,Fan:2022eto} and of the finestructure constant
	$\alpha$ \cite{Morel:2020dww,Parker:2018vye} is used, see also the discussion in the reviews
	\cite{Athron:2025ets,Keshavarzi:2022kpc}.
}

\section{Phenomenological analysis including flavour violating observables}
\label{sec:LFV}

Here we present a detailed analysis of the phenomenology of the six VLL models in the parameter regions motivated
by the new results for $a_\mu$ and $R_{\mu\mu}$ discussed above, with particular focus on the possible predictions for CLFV observables. We 
begin by setting up the relevant observables, then we discuss the roles
of chiral enhancement and dipole dominance, and we illustrate the rich
interplay between flavour-violating and flavour-conserving constraints. Finally we analyse the possible predictions for future measurements of low-energy and high-energy collider observables.

\subsection{Observables}

We consider the following set of CLFV observables:
radiative lepton decays $e_i\to e_j\gamma$, three-body decays $e_i\to e_j e_k\bar{e}_l$, the $\mu\to e$  conversion rate in presence of a nucleus, 
as well as bosonic decays $Z\to e_i\bar e_j$ and $h\to e_i\bar e_j$.
Except for the $Z$ decays, all of these observables receive chirally enhanced contributions similar to the 
mass corrections discussed in Sec.~\ref{sec:LFC}. In the leptonic decays these enter via the loop-induced electromagnetic form factors,
while in the Higgs decays they already appear at tree-level. The leptonic three-body decays and $\mu\to e$ conversion rate
additionally receive contributions from $Z$-exchange, that are not chirally enhanced but generated at tree-level rather than one-loop.
In the following we provide the analytical formulas in terms
of the effective couplings defined in Sec.~\ref{sec:eff-cpl}. For the evaluation we perform the exact numerical diagonalization and compute
the couplings via the expressions listed in App.~\ref{App:couplings}. \\

\begin{table}
	\centering
	\def\arraystretch{1.2}
	\begin{tabular}{l|cc|cc}
		\textbf{Observable} & \multicolumn{2}{c|}{\textbf{current bound}}  & \multicolumn{2}{c}{\textbf{future sensitivities}} \\ \hline\hline
		BR$(\mu\to e\gamma)$ & $\SI{3.1e-13}{}$ & \cite{MEGII:2023ltw}&$6\times 10^{-14}$ & \cite{MEGII:2018kmf} \\
		BR$(\tau\to e\gamma)$ & $\SI{3.3e-8}{}$ & \cite{BaBar:2009hkt} &$ \SI{3e-9}{}$ & \cite{Belle-II:2018jsg}\\
		BR$(\tau\to \mu\gamma)$ & $\SI{4.2e-8}{}$ & \cite{Belle:2025bpu}& $\SI{1e-9}{}$ & \cite{Belle-II:2018jsg}  \\	\hline	
		BR$(\mu\to  3e)$ & $\SI{1e-12}{}$ & \cite{SINDRUM:1987nra} & 
		$\SI{1e-16}{}$ & \cite{Hesketh:2022wgw} \\
		BR$(\tau\to 3e)$ & $\SI{2.7e-8}{}$ & \cite{Hayasaka:2010np}& $\SI{4.6e-10}{}$ & \cite{Belle-II:2018jsg} \\
		BR$(\tau\to 3\mu)$ & $\SI{1.9e-8}{}$ & \cite{Belle-II:2024sce}& $\SI{3.6e-10}{}$ & \cite{Belle-II:2018jsg} \\
		BR$(\tau\to \mu\, 2e)$ &$\SI{1.8e-8}{}$ & \cite{Hayasaka:2010np} & $\SI{3e-10}{}$ & \cite{Belle-II:2018jsg} \\
		BR$(\tau\to e\, 2\mu)$ & $\SI{2.7e-8}{}$ & \cite{Hayasaka:2010np}&  $\SI{4.6e-10}{}$ & \cite{Belle-II:2018jsg} \\ \hline
		BR$_\text{Au}(\mu\to e)$ & $\SI{7e-13}{}$ & \cite{SINDRUMII:2006dvw}& --  & \\		
		BR$_\text{Al}(\mu\to e)$ & -- & & 
		 $3\times 10^{-17}$  & \cite{COMET:2018auw,Mu2e:2014fns}  \\ \hline
		BR$(Z\to\mu e)$    & $\SI{2.6e-7}{}$ & \cite{ParticleDataGroup:2024cfk} & $\SI{1e-10}{}$ & \cite{Dam:2018rfz} \\
		BR$(Z\to\tau e)$   & $\SI{5.0e-6}{}$ & \cite{ParticleDataGroup:2024cfk} & $\SI{1e-9}{}$ & \cite{Dam:2018rfz} \\
		BR$(Z\to\tau \mu)$ & $\SI{6.5e-6}{}$ & \cite{ParticleDataGroup:2024cfk} & $\SI{1e-9}{}$ & \cite{Dam:2018rfz} \\
		BR$(h\to\mu e)$    & $\SI{4.4e-5}{}$ & \cite{CMS:2023pte} & $\SI{1.2e-5}{}$ & \cite{Qin:2017aju} \\
		BR$(h\to\tau e)$   & $\SI{2.0e-3}{}$ & \cite{ATLAS:2023mvd} & $\SI{1.6e-4}{}$ & \cite{Qin:2017aju}\\
		BR$(h\to\tau \mu)$ & $\SI{1.5e-3}{}$ & \cite{CMS:2021rsq} & $\SI{1.4e-4}{}$ & \cite{Qin:2017aju}\\
	\end{tabular}
	\caption{List of current bounds and future sensitivities of flavour-violating observables.}
	\label{tab:LFVbounds}
\end{table}

The branching ratios of radiative decays of the form $e_i \to e_j\gamma $ are given by
\begin{align}\label{eq:ltolgamma}
	\text{BR}(e_i\to e_j \gamma) = \frac{\alpha m_i}{2\Gamma_i} \Big(|F_M^{ji}(0)|^2 + |F_D^{ji}(0)|^2\Big)
\end{align}
in terms of the dipole form factors $F_{M,D}(0)$ in Eq.~\eqref{eq:cov-dec-photon-vertex}.

For the three-body decays we adapt the results from Ref.\ \cite{Crivellin:2013hpa}, using essentially the same notation as the original authors and write the branching ratio as
\begin{align}
	\text{BR}(e_i\to e_j e_k\bar e_l) &= \frac{m_i^5 S_{jk}}{6144\pi^3 \Gamma_i} \bigg\{ \sum_{XY} \Big(4 |C^V_{XY}|^2 + |C^S_{XY}|^2\Big) + 
	48\Big(|C^T_L|^2 + |C^T_R|^2\Big) + X_\gamma  \bigg\}.
\end{align}
Here the symmetry factor $S_{jk} = \frac{1}{2}$ if $j=k$ and $1$ otherwise. 
Similarly, the photon contributions $X_\gamma$ depend 
on the final state. For the cases $(A): j=k=l$, $(B):k=l\neq j$ and $(C):$ else, $X_\gamma$
is given by
\begin{align*}
	X_\gamma^{(A)} &= \frac{64e^2}{m_i^2} \sum_X \Bigg\{\bigg[\ln(\frac{m_i^2}{m_j^2})- \frac{11}{4}\bigg] |C^\gamma_X|^2
	- \frac{m_i}{4e} \Re\bigg[\Big(2C^V_{XX}+C^V_{X\bar{X}}-\tfrac{1}{2}C^S_{X\bar{X}}\Big) C^\gamma_{\bar{X}}\bigg]  \Bigg\}, \\
	X_\gamma^{(B)} &= \frac{32e^2}{m_i^2} \sum_X \Bigg\{\bigg[\ln(\frac{m_i^2}{m_j^2}) - ~3\,\,\bigg] |C^\gamma_X|^2
	- \frac{m_i}{2e} \Re\bigg[\Big(C^V_{XX}+C^V_{X\bar{X}}\Big) C^\gamma_{\bar{X}}\bigg]  \Bigg\}, \\
	X^{(C)}_\gamma &= 0,
\end{align*}
where $\bar X= R,L$ for $X=L,R$. In these equations the CLFV form factors enter via
\begin{align}
	C_{L/R}^\gamma = \frac{e}{m_i+m_j} \Big(F_M^{ji}(0) \pm iF_D^{ji}(0)\Big).
\end{align}
The additional relevant contributions in our case stem from tree-level exchange of a $Z$-boson. 
In principle there are also contributions from tree-level
Higgs and Goldstone boson exchange, however, these are strongly suppressed by the small lepton masses and
can be ignored.
Nonetheless, coefficients of the scalar quadrilinears defined in Ref.\ \cite{Crivellin:2013hpa} are induced
after using Fierz identities \cite{Nishi:2004st,Isidori:2023pyp} to bring the entire amplitude into the
specified form. We obtain the non-vanishing coefficients ($X,Y \in \{L,R\}$)
\begin{align}
	C_{XX}^V = -\frac{1}{M_Z^2} \Big(g^{Ze}_{X,ji} g^{Ze}_{X,kl} + g^{Ze}_{X,ki} g^{Ze}_{X,jl}\Big), \quad
	C_{XY}^V = -\frac{1}{M_Z^2} g^{Ze}_{X,ji} g^{Ze}_{Y,kl}, \quad
	C_{XY}^S = +\frac{2}{M_Z^2} g^{Ze}_{Y,ki} g^{Ze}_{X,jl}.
\end{align}

For the $\mu\to e$ conversion process in presence of a nucleus $N$ we
follow the setup and notation of Ref.~\cite{Kitano:2002mt}.
The conversion rate is given by
\begin{align}
	\text{BR}_{N}(\mu\to e) &= \frac{2 G_F^2 m_\mu^5}{\Gamma_{\text{capt}}} \bigg\{ \Big| A_R^* D + \tilde g_{LS}^{(p)} S^{(p)} + \tilde g_{LS}^{(n)} S^{(n)} + 
	\tilde g_{LV}^{(p)} V^{(p)} + \tilde g_{LV}^{(n)} V^{(n)} \Big|^2 + L \leftrightarrow R  \bigg\}.
\end{align}
Here the capture rates $\Gamma_{\text{capt}}$ and the numerical
overlap integrals $D, S^{(i)}, V^{(i)}$ are  listed  in
Tab.~\ref{tab:overlap-int} for the two relevant cases $N=$Al, Au. The
normalization  of the overlap integrals differs from the one in
Ref.~\cite{Kitano:2002mt} by the dimensionful factor
$m_\mu^{5/2}$. 

\begin{table}
	\centering
	\def\arraystretch{1.2}
	\begin{tabular}{c|c|c|c|c|c|c}
		N & $D\  $ & $S^{(p)}$ & $S^{(n)}$ & $V^{(p)}$ & $V^{(n)}$ & $\Gamma_\text{capt}~[10^6\,\text{Hz}] $ \\ \hline\hline
		$\prescript{27}{13}{\text{Al}}$ & 0.0362 & 0.0155 & 0.0167 & 0.0161 & 0.0173 & 0.7054 \\ \hline 
		$\prescript{197}{79}{\text{Au}}$ & 0.189 & 0.0614 & 0.0918 & 0.0974 & 0.146 & 13.07 \\ 
	\end{tabular}
	\caption{Values of the $\mu\to e$ overlap integrals and capture rates for the relevant nuclei, taken from Ref.~\cite{Kitano:2002mt}}
	\label{tab:overlap-int}
\end{table}

The dipole operators contribute via the coefficients
\begin{align}
	A_{L/R}^* = -\frac{\sqrt{2} e}{8 G_F m_\mu^2} \Big( F_M^{12}(0) \pm i F_D^{12}(0) \Big).
\end{align}
The additional coefficients are linear coefficients of contributions from four-fermion operators and are given by
\begin{align}
	\tilde g_{XV}^{(p)} = 2 g_{XV(u)} + g_{XV(d)}, \qquad
	\tilde g_{XV}^{(n)} = g_{XV(u)} + 2 g_{XV(d)}.
\end{align}
The actual VLL contributions corresponding to four-fermion operators generated by tree-level $Z$-exchange are
\begin{align}
	g_{XV(q)} = \frac{\sqrt{2} e}{c_W s_W} \frac{g^{Ze}_{X,21} g^q_{V}}{G_F M_Z^2},
\end{align}
where again $X=L,R$ and $g_V^u = \frac{1}{4} - \frac{2}{3} s_W^2$, and $g_V^{d,s}=-\frac{1}{4}+\frac{1}{3} s_W^2$.
The scalar coefficients $g_{XS}^{(n,p)}$ induced by Higgs exchange, though implemented, are negligible.\footnote{%
As a remark on the precision of our computations we note that we take
into account only the lowest non-vanishing orders for all
observables. Especially for the $\mu\to e$ process we note that
improved precision requires treating the problem in a proper effective
field theory setup. Then renormalisation-group running from the BSM or
electroweak scale down to the nuclear scale establishes a more
accurate relationship between fundamental BSM parameters and the
effective low-energy dynamics, and improved computations of nuclear
and atomic effects are possible. For recent literature on these
improvements we refer to
Refs.~\cite{Crivellin:2017rmk,Haxton:2022piv,Haxton:2024lyc,Heinz:2024cwg,Noel:2024led,Fontes:2025mps}. As
an example we checked that using the elaborate setup of
Ref.~\cite{Haxton:2024lyc} would change our results by
$\mathcal{O}(10\%)$. For the purposes of the present paper of
surveying patterns of predictions of many observables in six models
this difference is negligible while the gain in evaluation speed of
using the setup of Ref.~\cite{Kitano:2002mt} is significant.
}

Finally, the CLFV bosonic branching ratios are given by
\begin{subequations}
	\begin{align}
		\text{BR}(h\to e_i \bar e_j) &= \frac{M_h}{8\pi\Gamma_h} \Big(\big|Y^{h}_{e,ij}\big|^2 + \big|Y^{h}_{e,ji}\big|^2\Big), \label{eq:LFV-Higgs-decay}\\
		\text{BR}(Z\to e_i \bar e_j) &= \frac{M_Z}{24\pi\Gamma_Z} \Big(\big|g^{Ze}_{L,ij}\big|^2 + \big|g^{Ze}_{R,ji}\big|^2\Big), \label{eq:LFV-Z-decay}
	\end{align}
\end{subequations}
where the experimental values of the Higgs and $Z$-boson decay widths are taken from \cite{ParticleDataGroup:2024cfk}.

\subsection{Chiral Enhancement and CLFV correlations}\label{sec:DD}

\begin{figure}[t]
	\centering
	\includegraphics[width=.29\textwidth]{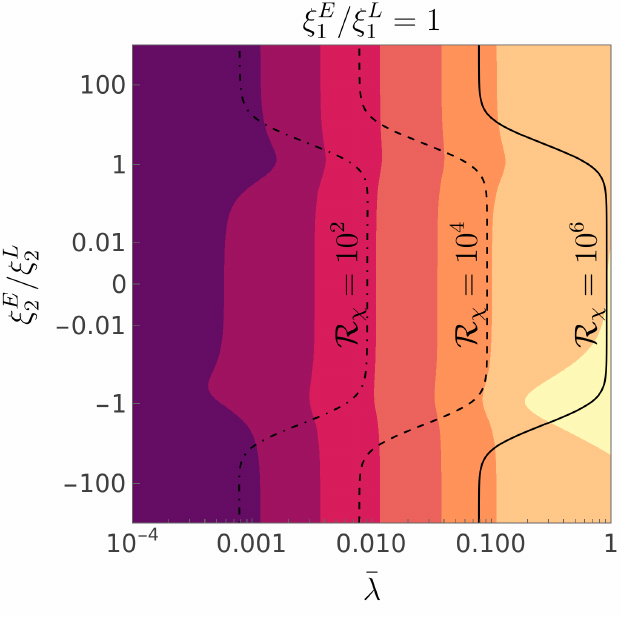}\hfill
	\includegraphics[width=.29\textwidth]{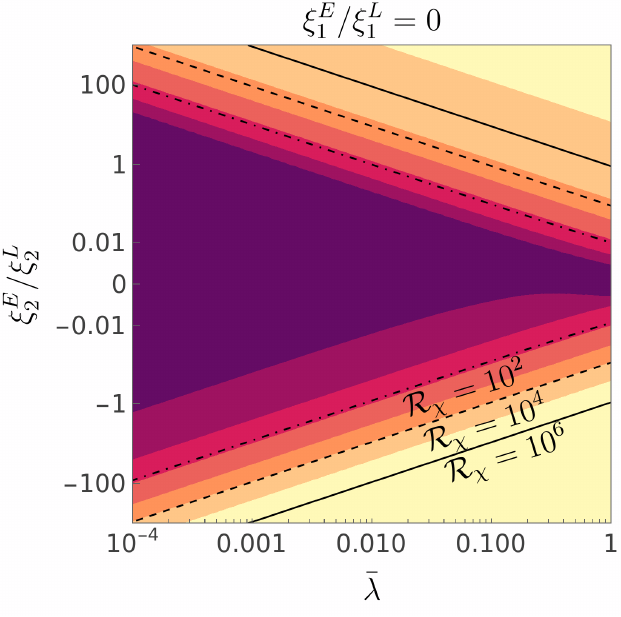}\hfill
	\includegraphics[width=.29\textwidth]{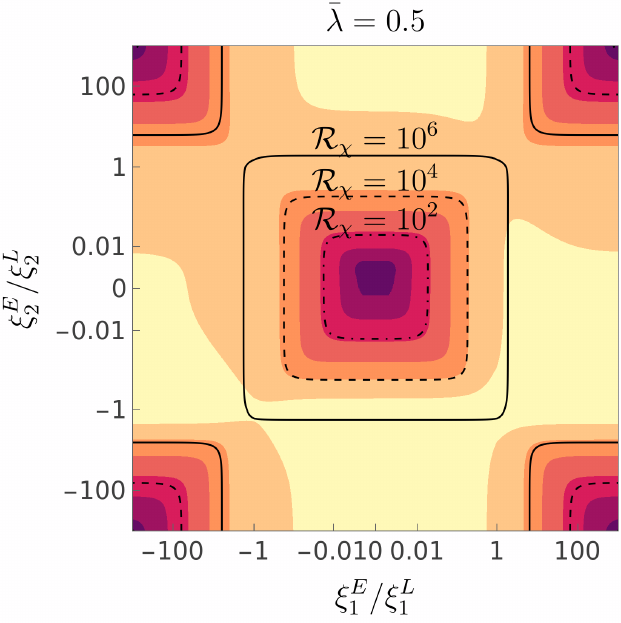}\hfill
	\raisebox{0.2\height}{\includegraphics[width=.07\textwidth]{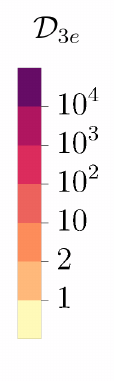}}
	\caption{Contours of $\mathcal{D}_{3e}^{21}$ (coloured) and $\mathcal{R}_{\chi}^{21}$ (black lines) for $\mu$--$e$ flavour violation 
		in the $L\oplus E$ model for different planes of the $\xi^E_1/\xi^L_1$ -- $\xi^E_2/\xi^L_2$ -- $\lambar$
		parameter space relevant to the CLFV ratios (assuming
                real couplings). Lighter colours correspond to regions
                with larger dipole contributions. Note that the
                $\xi^E/\xi^L$ axes are log-scaled, but include both signs.}
	\label{fig:DD-Rchi-plots}
\end{figure}

Similar to the discussion in Sec.~\ref{sec:LFC}, the presence of additional chiral symmetry breaking terms also leads to enhanced contributions in 
CLFV observables, induced by the off-diagonal lepton--Higgs couplings Eq.~\eqref{eq:Higgs-coupling} and electromagnetic form-factors Eq.~\eqref{eq:form-factors}. 
Here, however, there are qualitative differences in how these
contributions enter compared to the non-enhanced terms.
It is instructive to introduce and contrast the notions of chiral
enhancement and of dipole dominance. The simplest CLFV decays
$e_i\to e_j\gamma$ receive only  contributions from dipole form
factors, both with and without chiral enhancements. A useful measure
of chiral enhancement in $e_i\to e_j$ transitions, in analogy to the discussion in
Ref.~\cite{Athron:2025ets} for $a_\mu$, is thus
\begin{align}
	\mathcal{R}_{\chi}^{ij} = \frac{\text{BR}(e_i\to e_j \gamma)}{\text{BR}(e_i\to e_j \gamma)_{\lambar=0}}.
\end{align}
If no chiral enhancement is present, $\mathcal{R}_\chi^{ij}\approx1$;
if chiral enhancement is present, $\mathcal{R}_{\chi}^{ij}\gg 1$ and
$\mathcal{R}_{\chi}^{ij}$ is 
of the order $\lambar^2/y_i^2$, though it also depends on the
$\xi$'s. Fig.~\ref{fig:DD-Rchi-plots} displays $\mathcal{R}_\chi^{21}$
as black contours.

For other CLFV processes like\footnote{In the following we focus only on three-lepton final states with identical flavour, since
processes like $e_i\to e_j e_k\bar{e}_k$ are qualitatively very similar while more exotic processes like $e_i\to e_k e_k \bar{e}_j$ are very strongly suppressed.} 
$e_i\to 3 e_j$ and $\mu\to e$ we can define a suitable measure of
``dipole dominance'' (DD) by
\begin{subequations}\label{eq:DD-definitions}
	\begin{align}
		\mathcal{D}_{3e}^{ij}& = \mathcal{D}_0^{ij} \frac{\text{BR}(e_i\to 3e_j)}{\text{BR}(e_i\to e_j\gamma)}, \qquad \text{with} \qquad 
		\mathcal{D}_0^{{ij}} = \begin{cases}
			160 & \mu\to 3e,\\
			95 & \tau\to 3e,\\
			500 & \tau\to 3\mu,
		\end{cases}\\
		\mathcal{D}_\text{Al} &= 370\; \frac{\text{BR}_\text{Al}(\mu\to e)}{\text{BR}(\mu\to e\gamma)}.
	\end{align}
\end{subequations}
In case of DD, i.e.~if only dipole operators are relevant for these observables, there is
a strict correlation \cite{Lindner:2016bgg,Crivellin:2013hpa,Kitano:2002mt,Crivellin:2017rmk,Kuno:1999jp}, and the normalizations 
are chosen such that $\mathcal{D} \approx 1$ corresponds to
DD, while otherwise  typically
$\mathcal{D} \gg 1$. Interestingly, large
chiral enhancement does not guarantee  DD. The reason is that the 
$e_i\to 3 e_j$ and $\mu\to e$ observables receive non-dipole, and
non-chirally enhanced contributions, but these contributions are
already generated at tree-level via $Z$ exchange, while the chirally
enhanced contributions
only enter via the loop-suppressed dipole
coefficient. Fig.~\ref{fig:DD-Rchi-plots} displays
$\mathcal{D}_{3e}^{21}$ contours
via the coloured areas.

Further insight into the behaviour of these ratios can be obtained from the perturbative expressions given in Sec.~\ref{sec:eff-cpl}.
Firstly, a simple criterion for chiral enhancement and DD for the
purely leptonic decays (assuming
$\xi^E\sim\xi^L$) depends only on the value of $\lambar$ and reads
\begin{subequations}\label{eq:DD-naive-estimate}
	\begin{alignat}{5}
		\mathcal{R}_\chi^{ij} &\gg 1 \quad&&\Leftrightarrow\quad && |\lambar| \gg \frac{m_i}{v} \frac{|c_{L,R}|}{6 |c_{LR}|} && = \begin{cases}
			(0.15 ~\dots~ 5.7) \times 10^{-4} & i=\mu, \\
			(0.25 ~\dots~ 9.6) \times 10^{-3} & i=\tau,
		\end{cases}\\
		\mathcal{D}^{ij}_{3e} &\approx 1 \quad&&\Leftrightarrow\quad && |\lambar| \gg \frac{m_i}{v}\frac{\pi}{\alpha} \frac{|q_{L,R}|}{|c_{LR}|} && = \begin{cases}
			0.03 ~\dots~ 0.25 & i=\mu,\\
			0.48 ~\dots~ 4.3 & i=\tau.
		\end{cases}
	\end{alignat}
\end{subequations}
Here on the right the lower limits on $|\lambar|$ are evaluated
for the different models (except $L\oplus N$ that has
$c_{LR}=0$).
Combined with the discussion in Sec.~\ref{sec:vacuum} and
Sec.~\ref{sec:LFC} this suggests that DD is almost excluded, while chiral 
enhancement can still be readily present.

In case of chiral enhancement without DD a more accurate approximation is given by
\begin{subequations}
	\begin{align}
		\mathcal{R}_\chi^{ij} &\approx \frac{36v^2}{m_i^2} |\lambar|^2 \bigg(\frac{c_{LR}^2 (|\xi^E_i\xi^L_j|^2 + |\xi^E_j\xi^L_j|^2)}{c_L^2 |\xi^E_i\xi^E_j|^2 
			+ c_R^2 |\xi^L_i\xi^L_j|^2}\bigg) ,
		\\
		\mathcal{D}_{3e}^{ij} &\approx \frac{2\pi m_i^2}{3\alpha v^2} \frac{\mathcal{D}_0^{{ij}}}{|\lambar|^2}
		\bigg(\frac{ \tilde q_L^2 |\xi^E_i\xi^E_j|^2 + \tilde q_R^2 |\xi^L_i\xi^L_j|^2}{ c_{LR}^2 (|\xi^E_i\xi^L_j|^2 + |\xi^E_j\xi^L_j|^2)}\bigg),\label{DD3eapprox}
	\end{align}
\end{subequations}
where we have introduced the abbreviations $\tilde q_L^2 = 2 q_L^2 (1-4s_W^2+6s_W^4)$ and $\tilde q_R^2 = q_R^2 (1-4s_W^2+12s_W^4)$.
It is interesting to note that, while there are five independent couplings, these ratios (and more generally all the CLFV ratios discussed below) 
depend only on the three parameters $\xi^E_i/\xi^L_i$ and
$\xi^E_j/\xi^L_j$ and $|\lambar|$ (at the level of the approximations
of Sec.~\ref{sec:eff-cpl}).

The behaviour of chiral enhancement versus dipole dominance
is illustrated by Fig.~\ref{fig:DD-Rchi-plots}, which
shows contours for the $\mu$--$e$ flavour violating ratios
$\mathcal{R}_\chi^{21}$ and $\mathcal{D}_{3e}^{21}$ in different
planes 
of the relevant 3-dimensional parameter space. The left-most plot
illustrates the simple estimate Eq.~\eqref{eq:DD-naive-estimate}; for
$\xi^E_1/\xi^L_1\sim\pm1$ and $\xi^E_2/\xi^L_2\sim\pm1$,
 chiral enhancement is present for $\lambar\gtrsim10^{-4}$ and is
 indeed proportional to $|\lambar|^2$, whereas DD
is achieved only at very large values $|\lambar|\gtrsim1$. 
Notably, there is an asymmetry in $\pm\xi^E/\xi^L$ stemming from the relative sign between the dipole and $Z$-exchange contribution (in the full expression for $\mathcal{D}_{3e}^{ij}$),
that leads to either de- or constructive interference and consequently
slightly de- or increases the DD ratio. Fig.~\ref{fig:DD-Rchi-plots}
(left) and even more Fig.~\ref{fig:DD-Rchi-plots} (middle) also show
the behaviour for large hierarchies between the $\xi$'s. As also
visible in Eq.~(\ref{DD3eapprox}), large chiral enhancement is
achieved much more easily (or not at all) if the two ratios
scale oppositely (equally). This is even more apparent in the right
plot, where the dependence on both ratios is shown for fixed
$\lambar=0.5$.

For the discussion in the following section is it also useful to note
the expression for $\mathcal{D}_{3e}$ in the absence of chiral
enhancement (i.e.~for $\lambar=0$) and
specifically for $\mu$--$e$ flavour violation\footnote{In the
following, suppressed flavour indices $ij$ correspond always to
$\mu$--$e$ flavour violation.} 
\begin{subequations}\label{eq:DD-no-enhancement}
	\begin{alignat}{4}
		\mathcal{D}_{3e}\Big|_{\lambar=0} &\approx 
		1.7\times 10^6 &&\times &&\bigg(\frac{ \tilde q_L^2 |\xi^E_1\xi^E_2|^2 + \tilde q_R^2 |\xi^L_1\xi^L_2|^2}{c_L^2 |\xi^E_1\xi^E_2|^2
			+ c_R^2 |\xi^L_1\xi^L_2|^2}\bigg),  \\
			\intertext{together with the one for $\mathcal{D}_\text{Al}$,}
		\mathcal{D}_\text{Al}\Big|_{\lambar=0} &\approx
		2.7\times 10^7&&\times ~&&\bigg(\frac{q_L^2 |\xi^E_1\xi^E_2|^2
			+ q_R^2 |\xi^L_1\xi^L_2|^2}{c_L^2 |\xi^E_1\xi^E_2|^2
			+ c_R^2 |\xi^L_1\xi^L_2|^2}\bigg),
	\end{alignat}
\end{subequations}
where loop suppressed terms in the numerators are neglected.
Since they depend on the same coupling combinations, the ratios here
are bounded between $\tilde q_L^2 / c_L^2$ and $\tilde q_R^2 / c_R^2$
(or $q_L^2/c_L^2$ and $q_R^2/c_R^2$ in case of
$\mathcal{D}_\text{Al}$).
While the VLL models generally resemble the three-field models
discussed in Ref.~\cite{Athron:2025ets}, these boundary values correspond 
to the single-field limits, where only one of the VLL couples to the
SM leptons. The numerical single-field values for the different VLL representations
are listed in Tab.~\ref{tab:DD-single-field}; they differ by several
orders of magnitude. In the $L\oplus N$ model, in which both $q_L =
c_{LR} = 0$, the leading contributions vanish, leaving only
the (neglected) loop-suppressed terms in the numerator. This leads to the single-field
limit of $N$ to effectively coincide with the dipole dominance limit
of $\mathcal{D}\approx 1$.    

\begin{table}
	\centering
	\def\arraystretch{1.2}
	\begin{tabular}{c|c|c|c|c|c|c}
		\textbf{Field} & $E$ & $N$ & $L$ & $L_{\frac{3}{2}}$ & $E^a$ & $N^a$ \\ \hline\hline
		$10^{-6}\times\mathcal{D}_{3e}$      & 2.2 & $10^{-6}$ & 0.014 & 1.9 & 0.0016 & 0.42 \\
		$10^{-7}\times\mathcal{D}_\text{Al}$ & 4.5 & $10^{-7}$ & 0.033 & 4.2 & 0.0034 & 0.88\\
	\end{tabular}
	\caption{Single-field limits of $\mathcal{D}_{3e}$ and $\mathcal{D}_{Al}$ for the different VLL representations. }
	\label{tab:DD-single-field}	
\end{table}

Lastly, there are also relevant correlations between the leptonic and bosonic CLFV decays. In analogy to the discussion
in Sec.~\ref{sec:LFC}, the chirally enhanced contribution to the Higgs decay Eq.~\eqref{eq:LFV-Higgs-decay} is directly correlated with the 
radiative lepton decays induced by the form factors
\begin{align}
	\frac{\text{BR}(h\to e_i \bar{e}_j)}{\text{BR}(e_i\to e_j\gamma)} &\approx \frac{1}{\mathcal{Q}^2} \frac{M_h}{m_i} \frac{\Gamma_i}{\Gamma_h} 
	\frac{16\pi}{\omega_i\alpha}  \approx \frac{1}{\mathcal{Q}^2}
	\begin{cases}
		1 & i=\mu, \\
		1690 & i=\tau. 
	\end{cases}
\end{align}
Here it is worth pointing out that the strong increase of the ratio
for $\tau$ flavour-violating processes is driven by the scaling of the
leptonic decay fractions in the denominator, not by the CLFV Higgs decay width which is
(approximately) independent of the final state masses. 
The measurability of these ratios will be discussed in more detail the next section.
On the other hand, the CLFV $Z$ decays of Eq.~\eqref{eq:LFV-Z-decay} do not receive any chirally enhanced contributions at tree-level and,
in the absence of dipole dominance, are thus strongly correlated with e.g. $e_i\to 3 e_j$
\begin{align}\label{eq:Z-3l-correlation}
	\frac{\text{BR}(Z\to e_i\bar e_j)}{\text{BR}(e_i\to 3e_j)} \approx \frac{64\pi}{\alpha}
	\frac{M_Z^5}{m_i^5} \frac{\Gamma_i}{\Gamma_Z} c_W^2 s_W^2
	\bigg(
	\frac{q_L^2 |\xi^E_i\xi^E_j| + q_R^2 |\xi^L_i\xi^L_j|}{\tilde q_L^2 |\xi^E_i\xi^E_j| + \tilde q_R^2 |\xi^L_i\xi^L_j|}\bigg)
	= \begin{cases}
		0.34~\cdots~0.39 & i=\mu,\\
		1.9~\cdots~ 2.2 & i=\tau.
	\end{cases}
\end{align}

\subsection{Interplay of constraints}

\begin{figure}[t]
	\centering
	\begin{subfigure}[t]{0.321\textwidth}
		\centering
		\includegraphics[width=\textwidth]{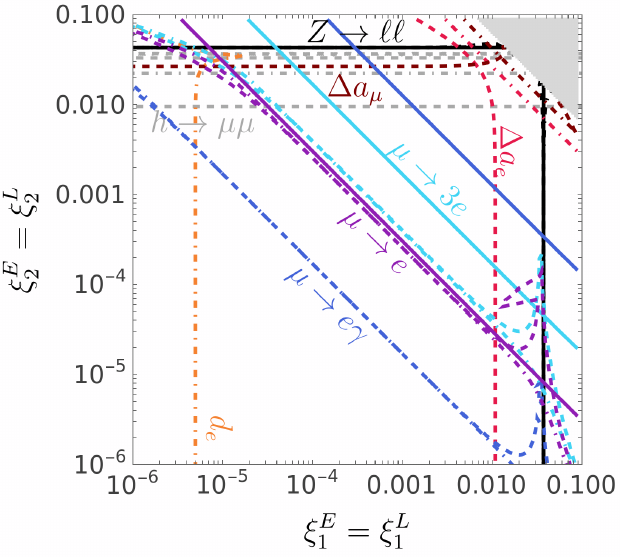}
		\caption{}
		\label{fig:interplaymue}\end{subfigure}
	\begin{subfigure}[t]{0.321\textwidth}
		\centering
		\includegraphics[width=\textwidth]{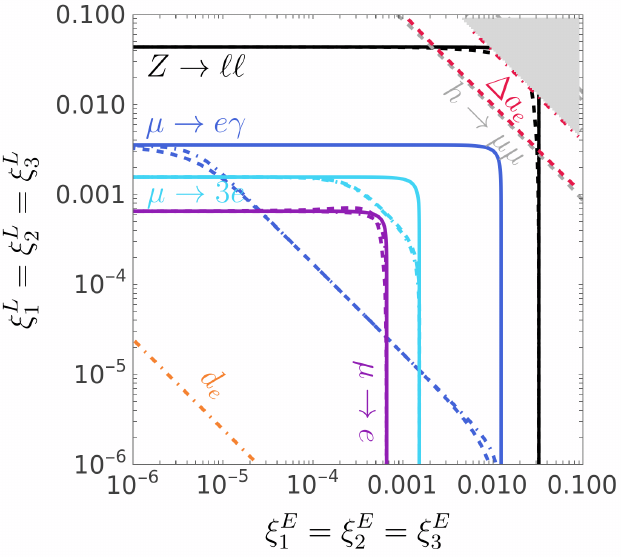}\hfill
		\caption{}
		\label{fig:interplayuniversal}\end{subfigure}
	\begin{subfigure}[t]{0.321\textwidth}
		\centering
		\includegraphics[width=\textwidth]{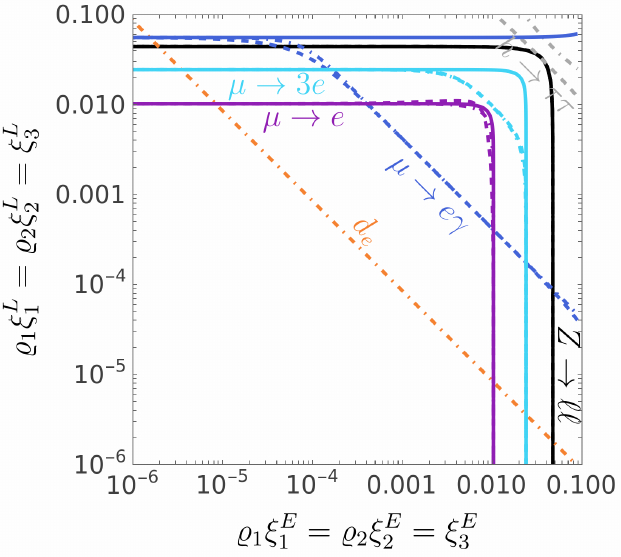}\hfill
		\caption{}\label{fig:interplaynonuniversal}
	\end{subfigure}
	\caption{$95\%$ CL exclusion lines for several
          observables, for   $\lambar=0$ (solid), $\lambar=-0.5$ (dashed) and 
		$\lambar=-0.5i$ (dot-dashed). The allowed regions are always to the
          bottom/left. The Dirac mass parameters are fixed at $2$ TeV and other parameters not shown are set to $0$. 
          In the third plot, $\rho_i = \sqrt{m_\tau / m_i}$.
		The grey region shows points with large precision loss in the SVD inversion	(i.e. relative error $> 10^{-5}$ compared to the input masses). }
    \label{fig:interplayconstraints}
\end{figure}

Here we analyse the interplay of the various flavour-conserving and
flavour-violating observables in constraining the VLL parameter
space. To be specific we pick the example of the $L\oplus E$ model,
but we note that the results for the other models would differ
essentially only because of the different $\O(1)$ coefficients in
Tab.~\ref{tab:cpl-coeff}.

The considered observables are the lepton--Z
couplings, the dipole moments $a_e$, $a_\mu$ and $d_e$, the Higgs
decays $h\to ee,\mu\mu,\tau\tau$, and the CLFV processes $\mu\to
e\gamma$, $\mu\to 3e$, and $\mu\to e $ conversion.
The essential model parameters are
\begin{align}
  \bar{\lambda},\
  \xi^{L}_i,\
  \xi^{E}_i\ (i=1,2,3).
\end{align}
All observables depend on the $\xi$'s, and most observables receive
potentially large chirally enhanced contributions
$\sim\bar\lambda$. Other model parameters do not enter in first
approximation.

In view of the structure of contributions,
Fig.~\ref{fig:interplayconstraints} shows three parameter planes: (a)
the $\xi_1$--$\xi_2$ plane focusing only on the electron--muon sector
and setting $\xi^L_i=\xi^E_i$; (b) the $\xi^E$--$\xi^L$-plane with
flavour-universal values $\xi_1=\xi_2=\xi_3$; (c) the $\xi^E$--$\xi^L$-plane with
non-universal values $\xi_i\propto m_i$. In each plot, exclusion
contours from all observables are shown for different values of
$\bar\lambda=0$ (solid), $\bar\lambda=-0.5$ (dashed),
and the complex value $\bar\lambda=-0.5i$ (dot-dashed).

The most basic constraint is the one from electroweak precision
observables, i.e.~from the lepton--$Z$ couplings collectively treated as
described before Fig.~\ref{fig:EWPO-excl}. As described there,
this constraint essentially amounts to individual limits on
$|\xi_i^E|$ and on $|\xi_i^L|$ with little dependence on
$\bar\lambda$. This is visible in all three plots by the approximately
vertical/horizontal black lines.

The magnetic dipole moments $a_e$ and $a_\mu$ are flavour and CP
conserving but chirality flipping observables. Without chiral enhancement, i.e.~for
$\bar\lambda=0$, the VLL contributions to these observables are
unmeasurably small, but for $\bar\lambda=-0.5$, $\Delta a_e$ leads to a
constraint on the product $\xi_1^L\xi_1^E$ and $\Delta a_\mu$ to a
constraint on $\xi_2^L\xi_2^E$. These constraints are visible as
vertical/horizontal red/brown dashed lines in Fig.~\ref{fig:interplaymue}
and the diagonal red dashed line in Fig.~\ref{fig:interplayuniversal}.
The constraints are invisible in Fig.~\ref{fig:interplaynonuniversal}
because of the chosen smaller values of $\xi_{1,2}$ in the plot.

For imaginary $\bar\lambda=-0.5i$, there is CP violation and the
electron electric dipole moment $d_e$ provides an important
constraint. Its behaviour is similar to the one from $\Delta a_e$ for $\bar\lambda=-0.5$, but
numerically it is even stronger as visible by the orange dot-dashed
lines.

The Higgs decays $h\to ee,\mu\mu,\tau\tau$ and the ratios
$R_{ee,\mu\mu,\tau\tau}$ are strongly correlated to the dipole moments
as discussed in Sec.~\ref{sec:LFC}. As a result, the constraint
$h\to\mu\mu$, shown in light grey in Fig.~\ref{fig:interplaymue} and Fig.~\ref{fig:interplayuniversal}, 
is similarly only relevant for sizeable $\bar\lambda$ but stronger than the one from $\Delta a_\mu$. 
The current bound on $h\to ee$ places no constraint. For large 3rd generation couplings, shown in Fig.~\ref{fig:interplaynonuniversal}, 
the constraint from $h\to\tau\tau$ becomes slightly stronger than $h\to\mu\mu$, but is weaker than the EWPO constraints.
In this case of Fig.~\ref{fig:interplaynonuniversal} there are also no competitive bounds from the magnetic dipole moments.

Finally we discuss the impact of the CLFV observables.
In Fig.~\ref{fig:interplaymue}, the CLFV constraints appear as essentially
diagonal lines. For $\bar\lambda=0$, the dipole observable $\mu\to e\gamma$ 
places only a very weak constraint, but $\mu\to3e$ and $\mu\to e$
significantly constrain the parameter space. For $\bar\lambda=-0.5$,
all three observables receive chirally enhanced dipole
contributions. The constraints from  $\mu\to3e$ and $\mu\to e$
strengthen mildly, but the one from $\mu\to e\gamma$ strengthens
substantially and becomes the strongest constraint. Since these observables are
insensitive to CP violation, $\lambar=-0.5i$ produces the same exclusion contours.
Notably, even in Fig.~\ref{fig:interplaynonuniversal} the $\mu$--$e$ CLFV bounds are still
significantly stronger than those from CLFV $\tau$-decay  observables.

\subsection{Characteristic Flavour Patterns}\label{sec:patterns}

\begin{figure}[t]
	\centering
	\includegraphics[width=.65\textwidth]{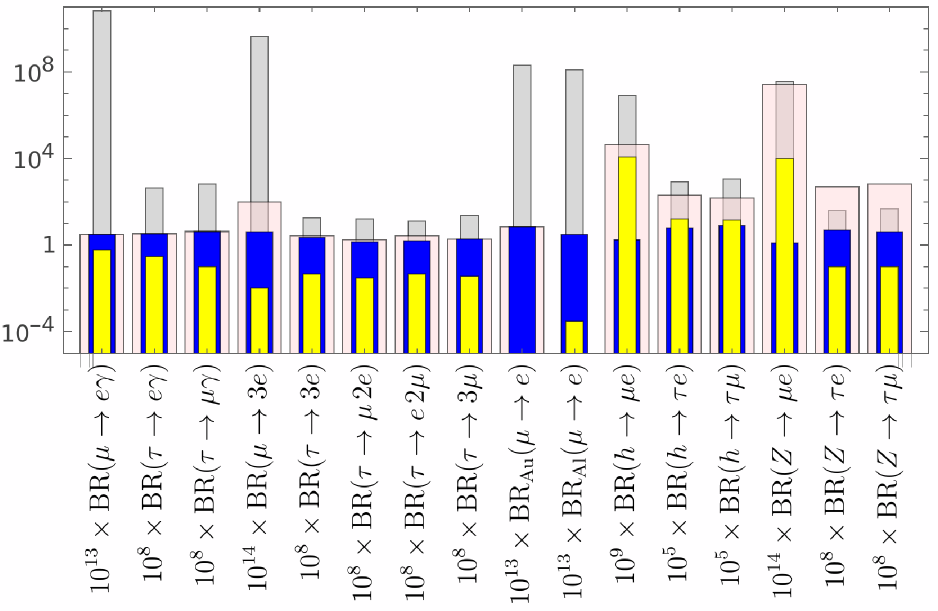}
	\raisebox{2\height}{\includegraphics[width=.2\textwidth]{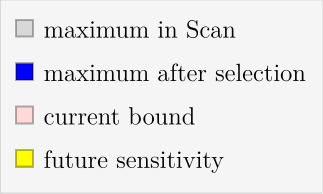}}
	\caption{Maximum ranges of CLFV observables in Scan II of the $L\oplus E$ model before (grey) and after selecting (blue) for the current
	experimental bounds. The current bounds and future sensitivities from Tab.~\ref{tab:LFVbounds} are shown in orange and yellow respectively.}
	\label{fig:LFV-ranges}
\end{figure}

Finally we analyse the possible predictions of all six VLL models for
observables of interest in future measurements. We focus both on
common features and on patterns of correlations that allow model
discrimination. The analysis is based on two random parameter scans for all six models (each $10^6$ points per model), using the
parameter ranges
\begin{itemize}
	\item \textbf{Scan I} (no CPV): $|\xi^{L,E}_i|\in (5
          \times10^{-11}, 0.05)$, $|\lambda|,|\lambar|\in(5\times
          10^{-11}, 0.25)$, $M_{L,E}\in (1~\text{TeV}, 5~\text{TeV})$
          with log-uniform sampling for the BSM Yukawa couplings and linear sampling for the masses
	\item \textbf{Scan II} (CPV): $|\xi^{L,E}_i|\in (5 \times10^{-9}, 0.05)$, $\arg(\xi^{L,E}_i)\in(0,2\pi)$, $|\lambda|,|\lambar|\in(-1,1)$, $M_{L,E} = 2~\text{TeV}$. Only the $\xi$'s are complex, with log-uniform sampling of the magnitude. All other parameters are real and sampled linearly.
\end{itemize}
Each of the following plots then shows resulting values of selected
observables, with colours corresponding to the six different
models. In the plots only points of the scans are used that satisfy
the current existing experimental constraints listed in Tab.~\ref{tab:LFCbounds} and Tab.~\ref{tab:LFVbounds}. 

Fig.~\ref{fig:LFV-ranges} provides an overview of the following
discussion by summarising the ranges of CLFV observables in 
Scan II\footnote{The analogous plots for Scan I as well as for the other models are qualitatively identical.} before and after the selection together
with the current and future sensitivities. By construction, the
maximum after selecting points that satisfy all current constraints
(blue) is equal to the current limits (light red) for a subset of
observables and smaller than the current limits for all other
observables. For all observables where the maximum after selection
(blue) is bigger than the future sensitivity (yellow), a detectable
signal at planned experiment is possible. This is the case in almost
all observables except the CLFV Higgs decays.

\begin{figure}[t]
	\centering
	\begin{subfigure}{.4\textwidth}
		\centering
		\includegraphics[width=\textwidth]{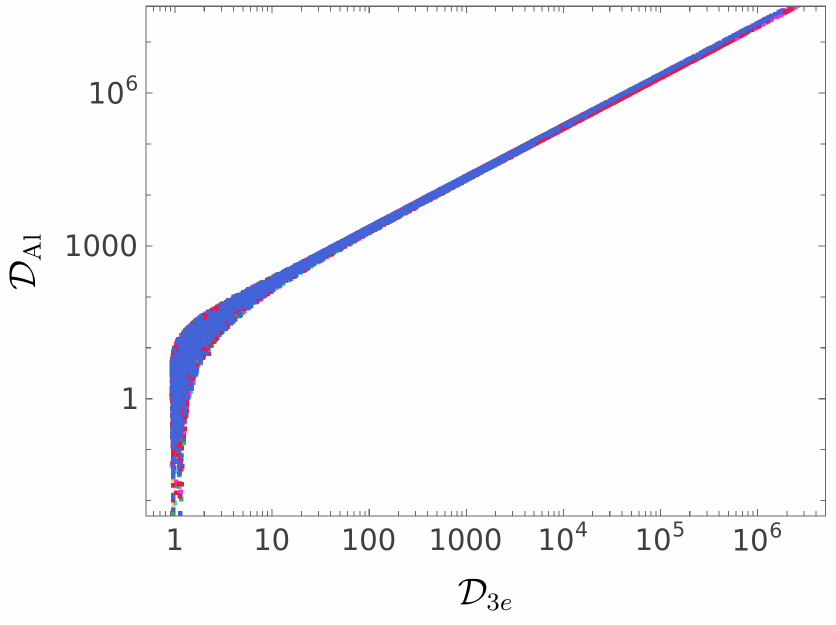}
		\caption{}
		\label{fig:DDvsDD}
	\end{subfigure}\hfill
	\begin{subfigure}{.4\textwidth}
		\centering
		\includegraphics[width=\textwidth]{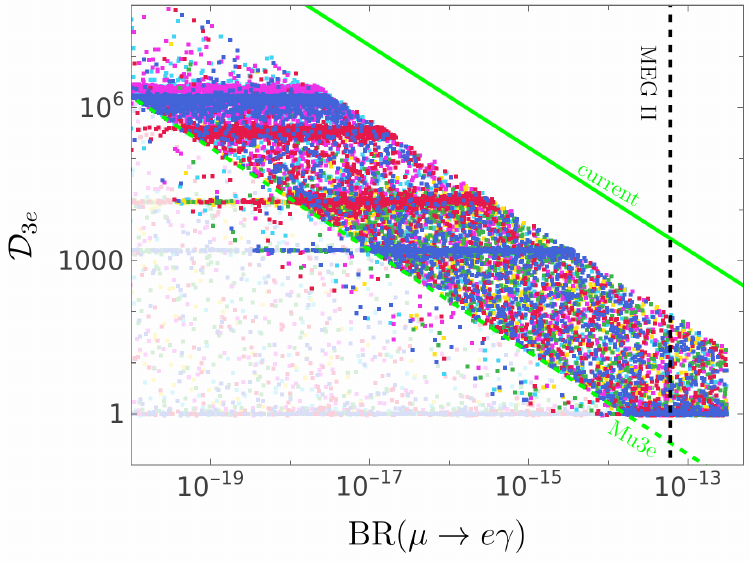}
		\caption{}
		\label{fig:DD3evsMueg}
	\end{subfigure}\hfill
	\begin{subfigure}{.1\textwidth}
		\centering
		\raisebox{.7\height}{\includegraphics[width=\textwidth]{Plots/legend.pdf}}
	\end{subfigure}
	\caption{\textbf{(a)} Correlation of dipole dominance ratios Eq.~\eqref{eq:DD-definitions} for points from Scan II.
	\textbf{(b)} Points from Scan I in the $\mathcal{D}_{3e}$--BR$(\mu\to e\gamma)$ plane.
	The dashed black line indicates the final MEG II sensitivity while the green (dashed) line indicates
	the current bound (and Mu3e sensitivity) for $\mu\to3e$.
	 The grey points are below the sensitivity of Mu2e/ COMET for $\mu\to e$ conversion in Al.}
	\label{fig:mueCLFV}
\end{figure}

Fig.~\ref{fig:DDvsDD}
shows the strong correlation between the two dipole dominance ratios $\mathcal{D}_{3e}$ and 
$\mathcal{D}_\text{Al}$, corresponding to the ratios between $\mu\to3e$ or $\mu\to e$
and $\mu\to e\gamma$ defined in Eq.~\eqref{eq:DD-definitions} and
discussed in Sec.~\ref{sec:DD}.
Far away from dipole dominance ($\mathcal{D}_i\gg 1$) Eq.~\eqref{eq:DD-no-enhancement} implies an upper and lower limiting value of the ratio $\mathcal{D}_{3e}/\mathcal{D}_\text{Al}$ that is independent of the coefficients $q_i$ and $\xi_i$, leading to 
the proportionality between the two ratios in all models visible in Fig.~\ref{fig:DDvsDD}. 
On the other hand, in the region close to dipole dominance
the interference between the dipole and $Z$ contribution weakens the correlation and
allows a larger range of values particularly for $\mathcal{D}_\text{Al}$.

Fig.~\ref{fig:DD3evsMueg} shows the range of possible $\mathcal{D}_{3e}$ as
a function of the $\mu\to e\gamma$ branching ratio. In view of Fig.~\ref{fig:DDvsDD},
the bound from BR$_\text{Au}(\mu\to e)$ behaves qualitatively similar to
the bound on BR$(\mu\to3e)$, but currently provides a stronger constraint and thus results
in the gap between the points and the green solid line. 
The plot also shows the sensitivity of the forthcoming MEGII and Mu3e experiments (dashed lines).

Fig.~\ref{fig:DD3evsMueg} illustrates that all six models can lead to
observable signals at Mu3e and possibly also at MEGII. While all models can
accommodate a wide range of values, in each model there are
``preferred'' values of $\mathcal{D}_{3e}$ that appear more frequently
in the scan. These correspond to the single-field limits listed in
Tab.~\ref{tab:DD-single-field}, and they are prominent in the plot
because of the non-negligible likelihood of large hierarchies between the couplings
due to the logarithmic sampling of $\xi_i^{L,E}$.

With these single-field limits in mind, several future scenarios can
be discussed. If signs of CLFV are observed in both $\mu \to e,
\gamma$ and $\mu \to 3e$ or $\mu \to e$ conversion the corresponding
ratio is fixed by observation, and we can distinguish the following outcomes. 
\begin{itemize}
	\item $\mathcal{D}_{i}$ is close to one. 
	In this case dipole dominance holds, and a plethora of models
        is able to explain the observation. For the VLL models this
        specifically needs large chiral enhancement implying at least
        moderately large $\lambar \gtrsim 0.01$, or dominance of the $N$ field in the $L\oplus N$ model. 
	\item $\mathcal{D}_{i}$ lies close to a value that can be explained by one of the single-field limits. This would be a hint towards the corresponding subset of models and the parameter region with very small or vanishing $\lambar$ and hierarchical values of $\xi^{L,E}$. 
	\item $\mathcal{D}_{i}$ lies somewhere in between the regions of single-field limits and DD.
	Both VLL fields would need to contribute significantly to explain the observation. As also illustrated by Fig.~\ref{fig:DD-Rchi-plots}, intermediate values of $\lambar$ can lead to such a pattern.
\end{itemize}
As Fig.~\ref{fig:DD3evsMueg} also clarifies, an improvement of the $\mu\to e\gamma$ sensitivity beyond the MEG II goal would be highly welcome to aid discriminating between the models.

\begin{figure}[t]
	\centering
	\begin{subfigure}{.4\textwidth}
		\centering
		\includegraphics[width=\textwidth]{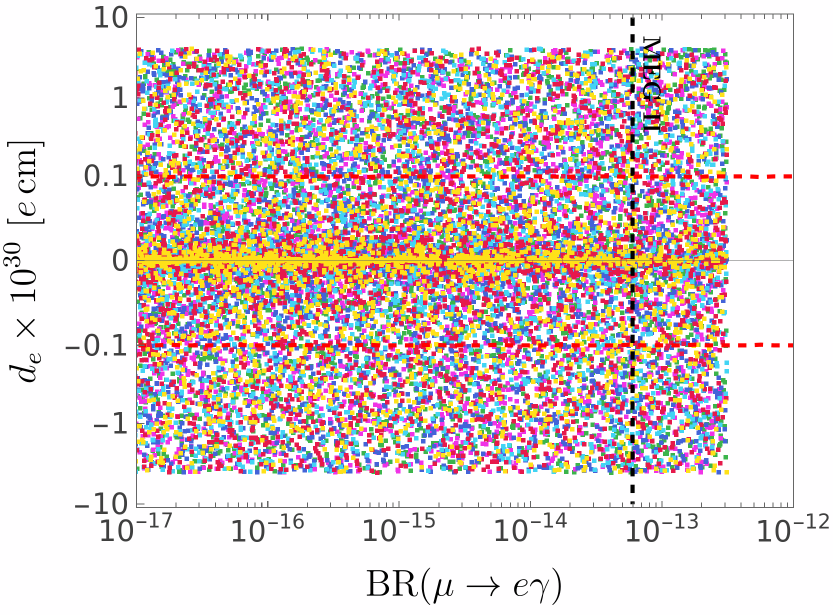}
		\caption{}
	\end{subfigure}\hfill
	\begin{subfigure}{.4\textwidth}
		\centering
		\includegraphics[width=\textwidth]{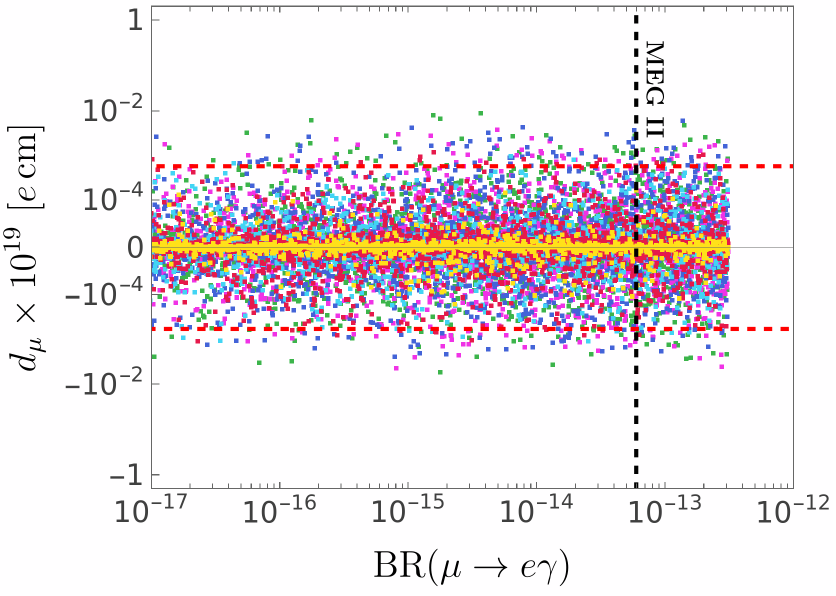}
		\caption{}
	\end{subfigure}\hfill
	\begin{subfigure}{.1\textwidth}
		\centering
		\raisebox{.7\height}{\includegraphics[width=\textwidth]{Plots/legend.pdf}}
	\end{subfigure}
	\caption{Points from Scan II in the BR$(\mu\to e\gamma)$ and \textbf{(a)} $d_e$ or \textbf{(b)} $d_\mu$ plane. The black dashed line indicates the final MEG II sensitivity and the
	red dashed lines the projected future sensitivities of $d_e$ and $d_\mu$ listed in Tab.~\ref{tab:LFCbounds}.}
	\label{fig:EDM}
\end{figure}

Similarly, Fig.~\ref{fig:EDM}
shows that the electron EDM $d_e$ can receive huge contributions
in all models, independently of the value of other observables such
as $\mu\to e\gamma$. The constraint on CP violation in the electron
sector is already strong at present and excludes parameter points with too large phases, and it will be further sharpened by
future progress. In contrast, the muon EDM $d_\mu$ cannot receive
contributions larger than $\SI{1e-21}{\ecm}$ in any of the
models. The reason is the correlation with
$h\to\mu\mu$ constrained by LHC, as already visible in the dipole
ellipse in Fig.~\ref{fig:dipoleellipses}.  

\begin{figure}[t]
	\centering
	\begin{subfigure}{.29\textwidth}
		\centering
		\includegraphics[width=\textwidth]{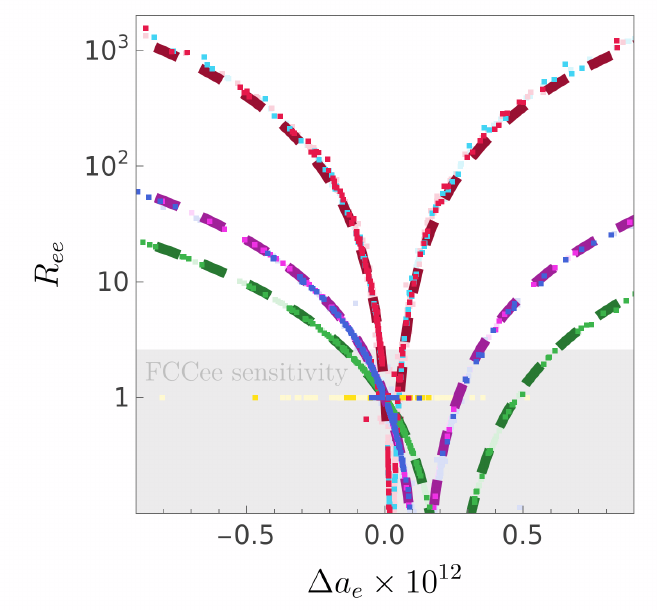}
		\caption{}
		\label{fig:Rllvsae}
	\end{subfigure}\hfill
	\begin{subfigure}{.29\textwidth}
		\centering
		\includegraphics[width=\textwidth]{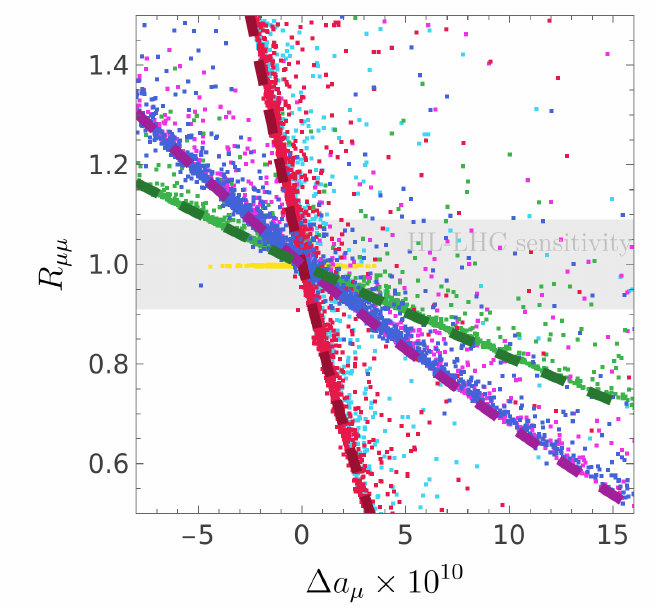}
		\caption{}
		\label{fig:Rllvsamu}
	\end{subfigure}
	\begin{subfigure}{.29\textwidth}
		\centering
		\includegraphics[width=\textwidth]{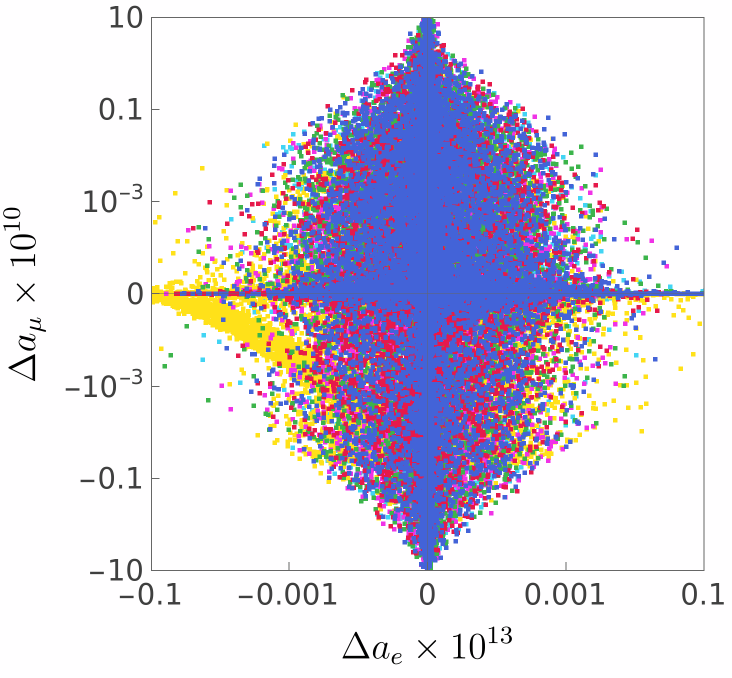}
		\caption{}
		\label{fig:ae-amu-MEGII}
	\end{subfigure}	\hfill
	\begin{subfigure}{.1\textwidth}
		\centering
		\raisebox{.7\height}{\includegraphics[width=\textwidth]{Plots/legend.pdf}}
	\end{subfigure}
	\caption{ \textbf{(a)} Correlation between $R_{ee}$ and $\Delta a_e$ in Scan I fulfilling all current bounds. 
		The greyed out points show points from Scan II with $|d_e|< 10^{-25}\SI{}{\ecm}$.
		\textbf{(b)} Correlation between $R_{\mu\mu}$ and $\Delta a_\mu$ for points in Scan II fulfilling all current bounds. 
		The grey bands indicate the projected FCC$ee$ ($R_{ee}$) and HL-LHC ($R_{\mu\mu}$) sensitivities sensitivity
		and the dashed lines correspond to the theory prediction Eq.~\eqref{eq:Rii-correlation} in case of CP conservation (i.e. $d_i=0$)
		with $\mathcal{Q}=1$ (cyan), $5$ (magenta) and $9$ (green).
		\textbf{(c)} Points visible in MEG II in the $\Delta a_e$ -- $\Delta a_\mu$ plane.}
	\label{fig:Rllvsae-amu}
\end{figure}

Figs.~\ref{fig:Rllvsae} and \ref{fig:Rllvsamu} again illustrate the correlation between the Higgs signal strength and the lepton dipole moments,
however, this time in the $\Delta a_i$ -- $R_{ii}$ plane.
In addition we also overlay the correlation predicted by Eq.~(\ref{eq:Rii-correlation}) in the case $d_i=0$ (dashed lines). 
For the electron we plot points from Scan I (coloured) with $d_e=0$ and Scan II (greyed-out)
with $|d_e|<10^{-25}\SI{}{\ecm}$. Both essentially perfectly follow the dashed lines, illustrating the strong constraint of 
the current bound on $d_e$ on CPV in the electron sector. We note that selecting points from Scan II that fulfil the actual
$d_e$ bounds also restricts $\Delta a_e \lesssim 10^{-15}$. The reason for this is that the linear sampling of the CPV phases in the scan
almost never produces tiny CPV phases of the $\xi_i$ required for a
relative suppression of $d_e$ compared to $\Delta a_e$, and thus points only evade the EDM bound if the overall scale of the $\xi_i$ 
is small enough.
In comparison, the bound on $d_\mu$ still allows for large CPV in the muon sector and the resulting deviations from the dashed lines
are correspondingly much larger.

The plots also show the sensitivities on $h\to ee$ and $h\to \mu\mu$ expected at the future
FCC$ee$ and high-luminosity LHC experiments. These sensitivities,
together with improved determinations of $\Delta a_{e,\mu}$ at the
level of $10^{-13}$ and $10^{-10}$, respectively, will be highly
beneficial for either identifying, discriminating or excluding the models. 
The correlations between the displayed observables essentially
depend only on the model-specific value of $\mathcal{Q}$ and are thus the
same for the pair $L\oplus E$ and $L\oplus N^a$ (cyan and pink), for
the pair  $L_{3/2}\oplus E$ and $L_{3/2}\oplus E^a$  (violet and
blue), while $L\oplus E^a$ (green) and $L\oplus N$ (yellow) each
predict a unique correlation.
Fig.~\ref{fig:ae-amu-MEGII} then shows the combined allowed values of
$\Delta a_e$ and $\Delta a_\mu$ visible within the MEG II sensitivity,
given the current experimental constraints. As expected, the CLFV 
bounds prohibit large values for both $\Delta a_{e,\mu}$ simultaneously
and instead prefer values significantly below the current sensitivities.

\begin{figure}[t]
	\centering
	\begin{subfigure}{.4\textwidth}
		\centering
		\includegraphics[width=\textwidth]{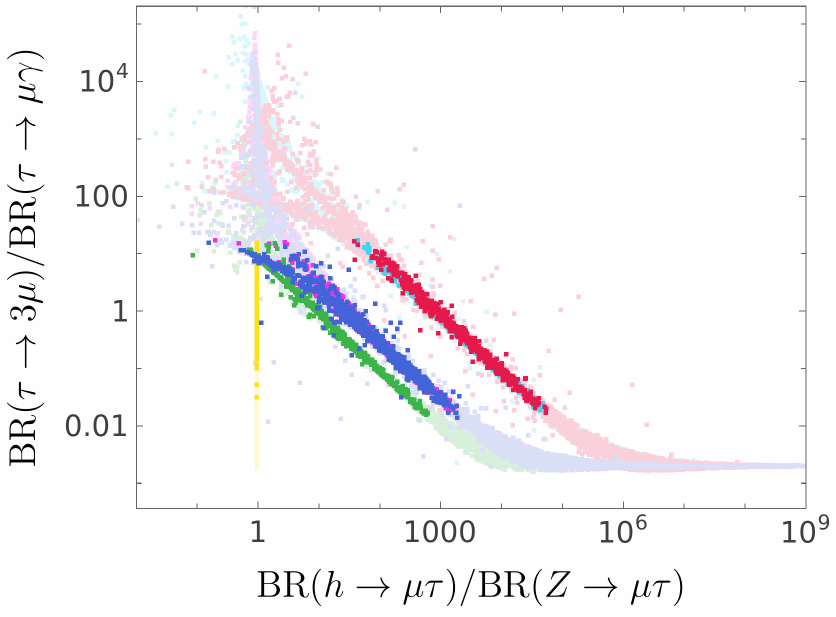}
		\caption{}
	\end{subfigure}\hfill
	\begin{subfigure}{.4\textwidth}
		\centering
		\includegraphics[width=\textwidth]{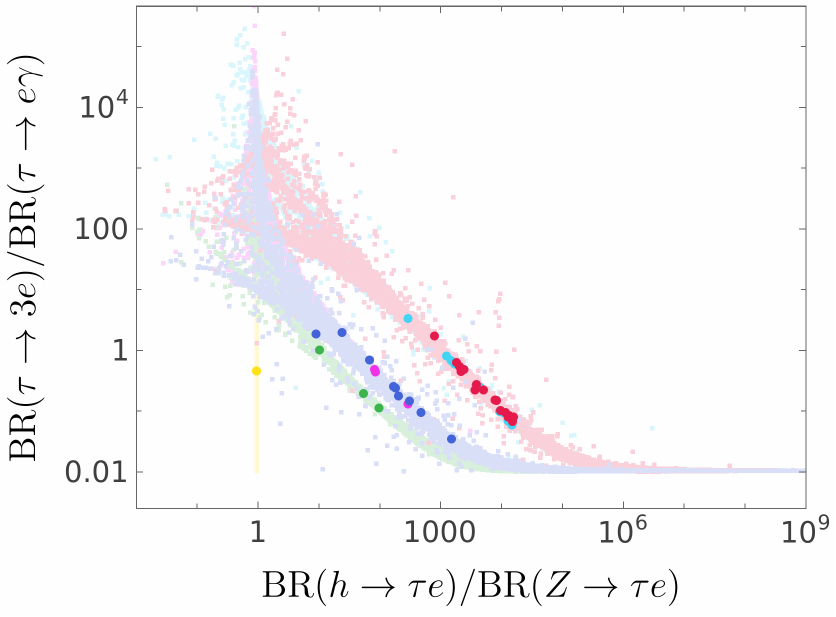}
		\caption{}
	\end{subfigure}
	\hfill
	\begin{subfigure}{.1\textwidth}
		\centering
		\raisebox{.7\height}{\includegraphics[width=\textwidth]{Plots/legend.pdf}}
	\end{subfigure}
	\caption{ \textbf{(a)} Correlations of CLFV observables with $\tau-\mu$ transitions (Scan II): the ratio of $\tau\to3\mu$ to $\tau\to\mu\gamma$ is plotted against the
		ratio of Higgs and $Z$ decays into $\mu\tau$. In the greyed out points either $\tau\to\mu\gamma$ lies below the sensitivity of Belle II
		or $Z\to\mu\tau$ below the sensitivity of the FCC$ee$.
		\textbf{(b)} analogous plot for $\tau-e$ CLFV. }
	\label{fig:FCC-vs-tau-decay}
\end{figure}

Next we include observables involving the $\tau$ lepton and $\tau$ flavour violation. As shown in Fig.~\ref{fig:LFV-ranges},
here also CLFV $Z$ (and to a lesser extent Higgs) decays become a promising probe in future experiments.
Fig.~\ref{fig:FCC-vs-tau-decay} shows the correlation between $\tau\to 3\ell$, $\tau\to\ell\gamma$
and the CLFV $Z$ and Higgs decays. The greyed out points are below the sensitivity of either Belle II or 
$Z\to \tau\ell$ at the FCC$ee$, while the coloured points are within
future sensitivities of these measurements.
The plots show that each model predicts a specific pattern of
correlations, i.e.~a thin possible band in the plane of ratios. As in
Fig.~\ref{fig:Rllvsae-amu}, the correlations are
the same for $L\oplus E$ and $L\oplus N^a$ (cyan and pink). For these
two models, the possible absolute values of the Higgs decay $h\to\tau\mu$ come
close to, but still lie slightly below the future sensitivity.
The $L_{3/2}\oplus E$ and $L_{3/2}\oplus E^a$ models (violet and blue)
also show an equal correlation with smaller values of $h\to\tau\mu$,
and the $L\oplus E^a$ (green) and $L\oplus N$ (yellow) models each
predict a unique correlation.

In case of $\tau$--$e$ transitions significantly fewer points yield detectable
values after selecting for the current bounds. The main reason is again the sampling of the CP
violating phases in Scan II; most points with
sizeable values of the couplings to electrons are excluded by the
constraint on $d_e$ shown in Fig.~\ref{fig:EDM}. In general, the plots
in Fig.~\ref{fig:FCC-vs-tau-decay} demonstrate that the correlations
between the observables are specific to the models; with a measurement
of all observables in the plots, at least a partial model discrimination would be in reach. 

\begin{figure}[t]
	\centering
	\begin{subfigure}{.4\textwidth}
		\centering
		\includegraphics[width=.49\textwidth]{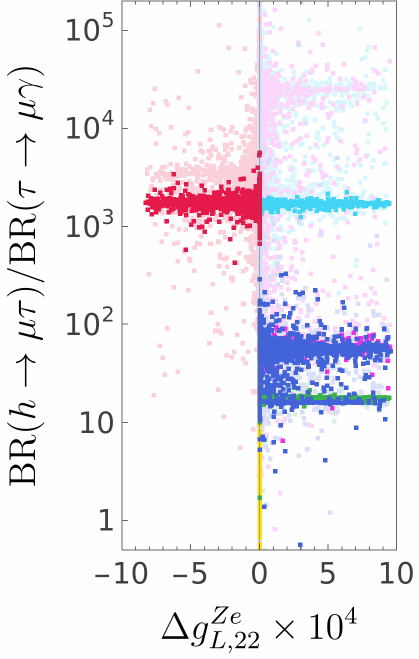}
		\includegraphics[width=.49\textwidth]{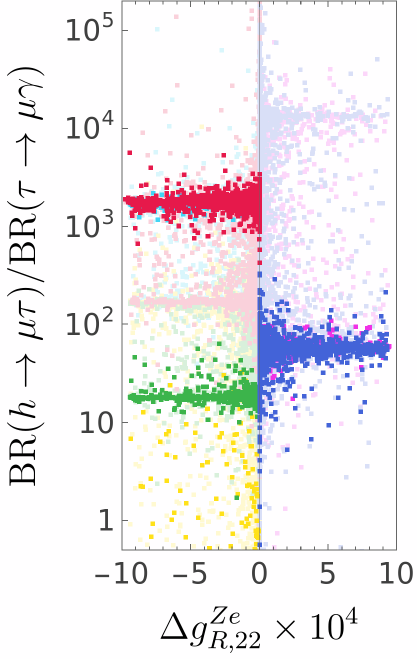}
		\caption{}
		\label{fig:tau-v-Higgs}
	\end{subfigure}\hfill
	\begin{subfigure}{.4\textwidth}
		\centering
		\includegraphics[width=.49\textwidth]{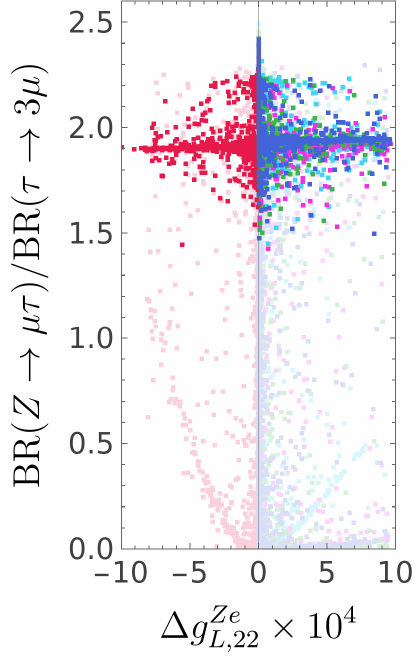}
		\includegraphics[width=.49\textwidth]{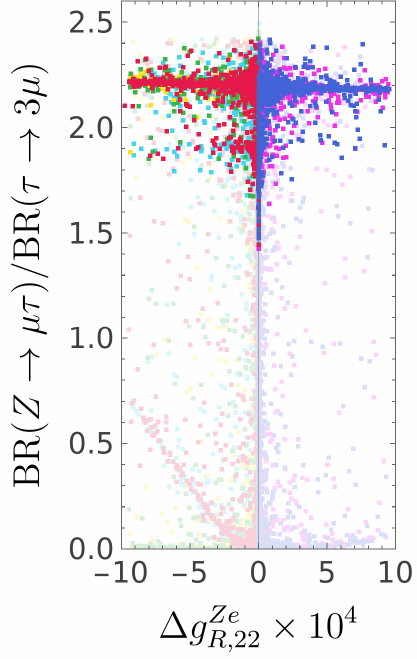}
		\caption{}
		\label{fig:tau-v-Z}
	\end{subfigure}\hfill
	\begin{subfigure}{.1\textwidth}
		\centering
		\raisebox{.7\height}{\includegraphics[width=\textwidth]{Plots/legend.pdf}}
	\end{subfigure}
	\caption{Scan II points for the ratio between \textbf{(a)} the $h\to\tau\mu$ and $\tau\to\mu\gamma$ decay widths
		and \textbf{(b)} $Z\to\tau\mu$ and $\tau\to 3\mu$ decay widths
		plotted against the VLL contribution to $g^{Ze}_{L,22}$ and $g^{Ze}_{R,22}$.
		The greyed out points are below the Belle II sensitivity for $\tau\to\mu\gamma,~3\mu$ or the FCC$ee$ sensitivity for $Z\to\tau\mu$.}
	\label{fig:tau-LVF-vs-EWPO}
\end{figure}

Fig.~\ref{fig:tau-LVF-vs-EWPO} again shows points that are visible at Belle II and (potentially) the FCC$ee$.
Here they are plotted against the VLL correction to the
flavour-conserving muon--$Z$ coupling. In Fig.~\ref{fig:tau-v-Higgs}
the $y$-axis displays the ratio of $h\to\mu\tau$ and
$\tau\to\mu\gamma$. Here the model
points show the clearest separation, i.e.~a measurement of these
observables would strongly aid in discriminating between the
models. In particular, the $L\oplus N^a$ model (red) is unique in
predicting negative corrections to the left-handed muon--$Z$
coupling. However, the appearing CLFV Higgs branching ratios are out of reach of the FCC$ee$.
On the other hand, Fig.~\ref{fig:tau-v-Z} shows the ratio
between $Z\to\tau\mu$ and $\tau\to3\mu$ on the $y$-axis, again against
the muon--$Z$ couplings. 
Importantly, here all four plotted
observables for the non-greyed points are within the foreseen
experimental sensitivities. For
these points in reach of future experiments the correlation essentially reduces to Eq.~\eqref{eq:Z-3l-correlation},
explaining the slightly weaker separation between the models.
Still, a
measurement of all the observables has the power to exclude a large
fraction of the VLL models.

\begin{figure}[t]
	\centering
	\begin{subfigure}{.21\textwidth}
		\includegraphics[width=\textwidth]{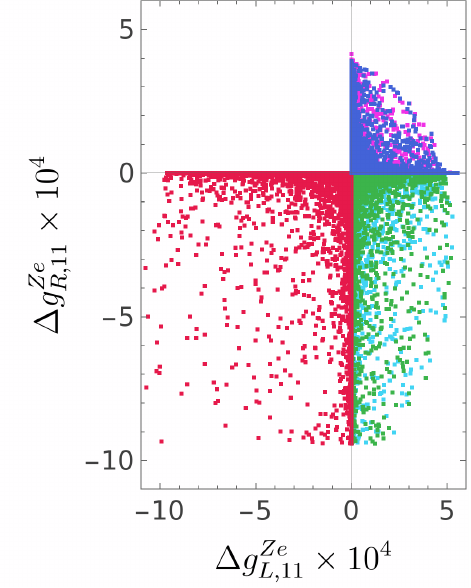}
		\caption{}
		\label{fig:gZR-gZL-e}
	\end{subfigure}\hfill
	\begin{subfigure}{.21\textwidth}
		\includegraphics[width=\textwidth]{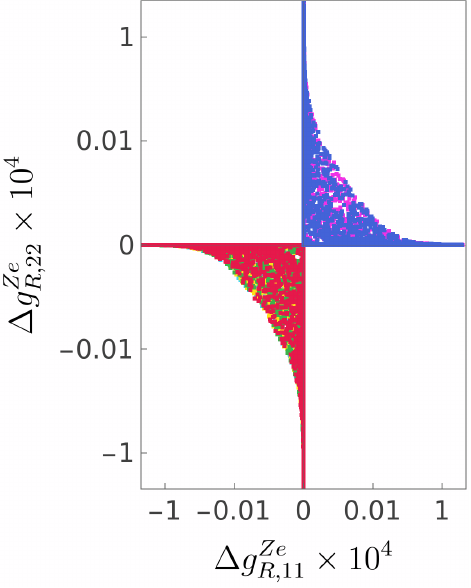}
		\caption{}
		\label{fig:gZR-mu-LFV}
	\end{subfigure}
	\begin{subfigure}{.21\textwidth}
		\includegraphics[width=\textwidth]{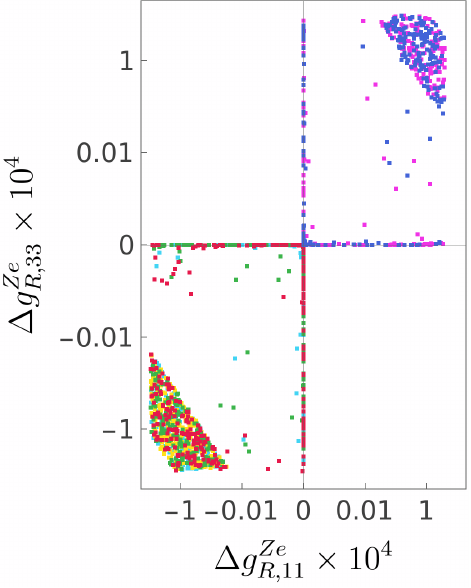}
		\caption{}
		\label{fig:gZR-tau-LFV}
	\end{subfigure}
	\begin{subfigure}{.21\textwidth}
		\includegraphics[width=\textwidth]{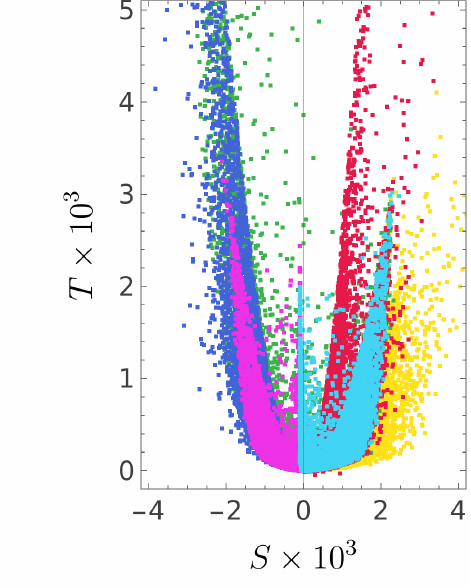}
		\caption{}
		\label{fig:S-T-correlation}
	\end{subfigure}
	\raisebox{.6\height}{\includegraphics[width=.1\textwidth]{Plots/legend.pdf}}
	\caption{Points from Scan I showing the correlation between VLL corrections to 
		\textbf{(a)} the left- and right-handed electron--$Z$ couplings and		
		\textbf{(b)} the right-handed electron- and muon--$Z$ couplings for points visible in Mu2e/COMET and Mu3e,
		\textbf{(c)} the right-handed electron- and tau--$Z$ couplings for points visible in Belle II,
		 \textbf{(d)} the electroweak oblique parameters $S$ or $T$. Only points are shown where at least one CLFV observable is within the sensitivities of Mu3e, Mu2e/COMET or Belle II}
	\label{fig:ST-vs-Zcpl}
\end{figure}

Finally, Fig.~\ref{fig:ST-vs-Zcpl} shows only parameter points that are within the sensitivities of either MEG II, Mu3e, Mu2e/COMET
or one of the CLFV measurements at Belle II. It thereby assumes the
existence of significant CLFV contributions, but the actual plots then
show the resulting viable ranges of flavour conserving
EWPO. Fig.~\ref{fig:gZR-gZL-e} plots the left- and right-handed
electron--$Z$ couplings and demonstrates a model separation at the level of 
$10^{-4}$ for these $Z$ couplings. The plot in
Fig.~\ref{fig:gZR-mu-LFV} shows the correlations between the right-handed $Z$ couplings
to electron and muon or to electron and $\tau$-lepton. Similar to the
dipole moments, the $\mu$--$e$ CLFV constraints already rule
out simultaneously large corrections to the electron-- and muon--$Z$ couplings. On the other hand, Fig.~\ref{fig:gZR-tau-LFV}
also shows that the CLFV in the $\tau$ sector prefers moderately large corrections to the $Z$
couplings to produce a visible CLFV signal.
The $S$ and $T$ parameters shown in Fig.~\ref{fig:S-T-correlation} may also aid in discriminating between the models
if the current sensitivity can be improved by two orders of magnitude, but are otherwise insensitive to CLFV.
In summary, the observables displayed in
Figs.~\ref{fig:Rllvsae-amu}--\ref{fig:ST-vs-Zcpl} are all
complementary for identifying or discriminating models.

\section{Conclusions and Outlook}

The six investigated VLL models lead to seesaw-like contributions to
lepton masses and thus change the mass generation mechanism, Higgs and
lepton phenomenology in ways that are complementary to many other
extensions of the SM. 
We have presented a broad phenomenological survey, taking into account
the implications of recent Higgs decay measurements and the new
determination of $a_\mu$. We have focused on collider observables
such as Higgs and $Z$  decays as well as low-energy lepton observables
such as $(g-2)$, EDMs and CLFV processes.

Each of the six models leads to a specific, strong correlation between
Higgs-to-lepton decays and lepton dipole moments. The recent
$h\to\mu\mu$ measurement then allows only narrow ellipses in the
$\Delta a_\mu$--$d_\mu$ plane --- or in case of CP conservation only
two specific regions for $\Delta a_\mu$ corresponding to normal-sign
and opposite-sign solutions for $\lambda_{\mu\mu}$. Combining this
result with the new constraint on $\Delta a_\mu$ leads to only small
viable regions of the dipole ellipses. These regions prefer values of
the BSM Yukawa coupling $\lambar$ that are moderate and not larger
than SM lepton Yukawa couplings (larger values are not entirely
excluded but more tightly constrained; their dedicated study is beyond
the scope of the present paper).

In this moderate-$\lambar$ parameter region, the models show a rich
lepton phenomenology. There typically are large chirally enhanced
contributions to Higgs couplings and lepton dipole operators, but
these do not necessarily lead to the familiar pattern of ``dipole
dominance'' for CLFV observables as shown in
Fig.~\ref{fig:DD-Rchi-plots}. Hence current results for $\mu\to
e\gamma$, $\mu\to 3e$, $\mu\to e$ conversion, $a_{e,\mu}$ and Higgs
and $Z$ decays all constrain the parameter space in relevant ways, as illustrated by Fig.~\ref{fig:interplayconstraints}.

In each model, the plausible ranges of predictions for CLFV
observables span many orders of magnitude, and observable signals in
all or a subset of the future MEGII, COMET, Mu2e and Mu3e experiments
are possible, see Fig.~\ref{fig:LFV-ranges}. However, each model leads to preferred ratios characterised by coupling hierarchies that are 
equivalent to single-field limits and appear more frequently as shown in Fig.~\ref{fig:mueCLFV}.

We have exhibited further model-specific  fingerprints,
i.e.~patterns of correlations between observables such as the magnetic
or electric dipole moments and Higgs signal strengths
(Figs.~\ref{fig:EDM} and \ref{fig:Rllvsae-amu}), correlations between
CLFV $\tau$ and $Z$ or Higgs decays (Fig.~\ref{fig:FCC-vs-tau-decay}),
as well as correlations of $Z$-lepton couplings with CLFV $\tau$ decays or electroweak oblique parameters (Figs.~\ref{fig:tau-LVF-vs-EWPO} and \ref{fig:ST-vs-Zcpl}). These can
allow future experiments to discriminate between the
models. Especially promising are observables including $\tau$-leptons,
such as CLFV $\tau$ decays measurable at Belle II and $Z$ decays
that could be measured e.g.~at FCC$ee$, but also future improvements for
electroweak precision observables will add highly valuable
information. Figs.~\ref{fig:tau-v-Z} and \ref{fig:ST-vs-Zcpl} 
show that the envisioned sensitivities are sufficient to distinguish
between or identify
several models in sizeable parameter regions. Even clearer model
separations would be possible if a higher sensitivity for CLFV Higgs
decays could be reached. Conversely, an observation of $h\to\tau\mu$
or $h\to\tau e$ at FCC$ee$ would exclude all six VLL models in all of
the considered parameter regions.

\section*{Acknowledgments}
 We acknowledge financial support by the German Research Foundation
 (DFG) under grant numbers STO 876/7-2 and STO 876/10-1.

\newpage
\appendix
\section{Mass-basis coupling matrices}\label{App:couplings}

\begin{table}[H]
	\centering
	\def\arraystretch{1.2}
	\begin{tabular}{|c|c|c|c|c|c|c|c|c|c|c|c|c|}
		\hline 
		\textbf{Field}
		& $N$ & $E$ & $L^0$ & $L^-$ & $L_{\frac{3}{2}}^-$ & $L_{\frac{3}{2}}^{--}$ & $N^+$ & $N^-$ & $N^0$ & $E^0$ & $E^-$ & $E^{--}$ \\ \hline \hline 
		$T^3$ & $0$ & $0$ & $\frac{1}{2}$ & $-\frac{1}{2}$ & $\frac{1}{2}$ & $-\frac{1}{2}$ & $1$ & $-1$ & $0$ & $1$ & $0$ & $-1$ \\ 
		$Y$ & $0$ & $-1$ & $-\frac{1}{2}$ & $-\frac{1}{2}$ & $-\frac{3}{2}$ & $-\frac{3}{2}$ & $0$ & $0$ & $0$ & $-1$ & $-1$ & $-1$ \\ 
		$Q$ & $0$ & $-1$ & $0$ & $-1$ & $-1$ & $-2$ & $1$ & $-1$ & $0$ &$0$ & $-1$ & $-2$ \\ \hline
	\end{tabular}
	\caption{Eigenvalues of $T^3, Y$ and $Q$ for the components of the VLL multiplets.}
	\label{tab:VLL-T3-Y-Q}
\end{table}

\begin{table}[H]
	\def\arraystretch{1.2}
	\begin{tabular}{|c||cc|}
		\hline
		$L$ & $L^0$ & $L^-$ \\ \hline\hline
		$L^0$ & 0 & $1$ \\
		$L^-$ & 0 & 0 \\\hline
	\end{tabular}\quad
	\begin{tabular}{|c||cc|}
		\hline
		$L_{\frac{3}{2}}$ & $L^-$ & $L^{--}$ \\ \hline\hline
		$L^-$ & 0 & $1$ \\
		$L^{--}$ & 0 & 0 \\\hline
	\end{tabular}\quad
	\begin{tabular}{|c||ccc|}
		\hline
		$E^a$ & $E^{--}$ & $E^-$ & $E^0$ \\ \hline\hline
		$E^{--}$ & 0 & 0 & 0 \\
		$E^-$ & $\sqrt2$ & 0 & 0 \\
		$E^0$ & 0& $-\sqrt2$ &0 \\ \hline
	\end{tabular}\quad
	\begin{tabular}{|c||ccc|}
		\hline
		$N^a$ & $N^{-}$ & $N^0$ & $N^+$ \\ \hline\hline
		$N^-$ & 0 & 0 & 0 \\
		$N^0$ & $\sqrt2$ & 0 & 0 \\
		$N^+$ & 0& $-\sqrt2$ &0 \\ \hline
	\end{tabular}
	\caption{Values of $\mathcal{G}^{Wff'}_{L,R}$ for the VLL doublets and triplets.}
	\label{tab:T+}
\end{table}

\begin{table}[H]
	\centering
	\def\arraystretch{1.2}
	\setlength{\tabcolsep}{3pt}
	\begin{tabular}{|c||cccc|}
		\hline
		$\mathcal{Y}^h_\nu$ & $N_R$ & $L_R^0$ & $N_R^0$ & $E_R^0$ \\ \hline\hline
		$\nu_{Li}$ & $\lambda^N_i$ & -- & $\lambda^N_i$ & $\sqrt2 \lambda^E_i$ \\
		$N_L$    & -- & $\lambar$ & -- & --  \\
		$L_L^0$  & $\lambda$ & -- & $\lambda$ & $\sqrt2\lambda$  \\
		$N_L^0$ & -- & $\lambar$ & -- & -- \\
		$E_L^0$ & -- & $\sqrt2\,\lambar$ & -- & -- \\ \hline
	\end{tabular}\hfill
	\begin{tabular}{|c||cccccc|}
		\hline
		$\mathcal{Y}^h_e$ & $ e_{Rj} $ & $E_R$ & $L_R^-$ & $L_{\frac{3}{2}R}^-$ & $N_R^-$ & $E^-_R$ \\ \hline\hline
		$e_{Li}$ & $y_i\delta_{ij}$ & $\lambda^E_i$ & -- & -- & $\sqrt2\lambda^N_i$ & $-\lambda^E_i$ \\
		$E_L$    & -- & -- & $\lambar$ &  $\lambar$ & -- & -- \\
		$L_L^-$  & $\lambda^L_j$ & $\lambda$ & -- & -- & $\sqrt2\lambda$ & $-\lambda$ \\
		$L_{\frac{3}{2}L}^-$ & $\lambda^L_j$ & $\lambda$ & -- & -- & -- & $\lambda$ \\
		$N_L^-$ & -- & -- & $\sqrt2 \,\lambar$ & -- & -- & --\\
		$E_L^-$ & -- & -- & $-\lambar$ & $\lambar$ & -- & -- \\ \hline
	\end{tabular}\hfill
	\begin{tabular}{|c||cc|}
		\hline
		$\mathcal{Y}^h_\rho$ & $L_{\frac{3}{2}R}^{--}$ & $E_R^{--}$ \\ \hline\hline
		$L_{\frac{3}{2}L}^{--}$ & -- & $\sqrt2 \lambda$  \\ 
		$E_L^{--}$ & $\sqrt2\,\lambar$ & --  \\ \hline
	\end{tabular}
	\caption{Yukawa couplings between the left- and right-handed lepton multiplet components}
	\label{tab:Yukawa-entries}
\end{table}

Here we collect the expressions for the mass-basis coupling matrices in the different charged seesaw models introduced in
Eqs.~\eqref{eq:L-MB-Yukawa} and \eqref{eq:L-MB-gauge}. The physical coupling matrices are given by ($X=L,R$)
\begin{subequations}
	\begin{align}
		&Y_f^h = U_L^{f\dagger} \mathcal{Y}_f U_R^f, \qquad  
		g_X^{Zf} =\tfrac{g_W}{c_W}U_X^{f\dagger}\mathcal{G}_X^{Zf} U_X^{f}, \qquad 
		g_X^{Wff'} =\tfrac{g_W}{\sqrt{2}}U_X^{f\dagger}\mathcal{G}_X^{Wff'} U_X^{f'}.
	\end{align}
\end{subequations}
The Goldstone-boson Yukawa matrices are related to the gauge couplings by \cite{Lynch:2001zs}
\begin{subequations}
	\begin{align}\label{eq:GB-cpl-unitariy}
		Y_{f}^{\phi^0} &= \frac{\sqrt2}{M_Z} \Big(m_f g^{Zf}_R - g^{Zf}_L m_f\Big), \\
		Y_{ff'}^{\phi^+} &= \frac{1}{M_W} \Big(g^{Wff'}_{L}m_{f'} - m_f g^{Wff'}_{R}\Big) \\
		Y_{ff'}^{\phi^-} &= \frac{1}{M_W} \Big(m_f g^{Wf'f\dagger}_R - g^{Wf'f\dagger}_{L} m_{f'}\Big)
	\end{align}
\end{subequations}
The matrices $\mathcal{Y}_f$, $\mathcal{G}_X^{Zf}$ and $\mathcal{G}_X^{Wff'}$ correspond to the couplings between the
multiplet components of the gauge-eigenstates. The gauge-boson couplings are given by the usual expression
\begin{align}
	\mathcal{G}_X^{Zf} = T^3 - Q s_W^2, \qquad
	\mathcal{G}_X^{Wff'} = T^+
\end{align}
where the eigenvalues of $T^3$ and $Q$ for the different VLL multiplet components are listed in Tab.~\ref{tab:VLL-T3-Y-Q}, and
the values of $T^+ = T^1 + i T^2$ for the different multiplets in Tab.~\ref{tab:T+}.
For the Yukawa coupling the entries are listed in Tab.~\ref{tab:Yukawa-entries}.\\

As an example how to reconstruct the explicit matrices for the VLL models, we consider $L\oplus N$.
The multiplet components mixing to form the mass-basis eigenstates are given in Eq.~\eqref{eq:mass-matrices},
\begin{align}
	U_L^{e\dagger} \hat e_{La} = \begin{pmatrix} e_{Li} \\ L_L^- \end{pmatrix}, \quad 
	U_R^{e\dagger} \hat e_{Ra} = \begin{pmatrix} e_{Rj} \\ L_R^- \end{pmatrix}, \quad
	U_L^{\nu\dagger}\hat\nu_{La} = \begin{pmatrix} \nu_{Li}\\ L_L^0 \\ N_L\end{pmatrix}, \quad 
	U_R^{\nu\dagger}\hat\nu_{Ra} = \begin{pmatrix} 0 \\ L_R^0 \\ N_R\end{pmatrix}.
\end{align}
The entries in the corresponding Yukawa matrices can then directly be read off from Tab.~\ref{tab:Yukawa-entries},
\begin{alignat}{4}
	&\left(\mathcal{Y}^h_{e}\right)_{L\oplus N} & &= \bordermatrix{ & e_{Rj} & L^-_R \cr 
				\overline{e_{L}}_{i} & y_{ij} & 0 \cr 
				\overline{L_L^-} & \lambda_j^L & 0 \cr } & \qquad
		&\left(\mathcal{Y}^h_\nu\right)_{L\oplus N}& &= \bordermatrix{ & 0 & L^0_R & N_R \cr 
				\overline{\nu_{L}}_{i} & 0 & 0 & \lambda^N_i \cr 
				\overline{L_L^0} & 0 &  0 & \lambda \cr 
				\overline{N_L} & 0 & \lambar & 0}.
\end{alignat}
The diagonal $Z$ coupling matrices are obtained from the coefficients\footnote{The SM lepton singlet and doublet 
	eigenvalues are equal to the VLL singlet and doublet $E$ and $L$.} in Tab.~\ref{tab:VLL-T3-Y-Q},
\begin{subequations}
	\begin{alignat}{3}
		&\left(\mathcal{G}_L^{Z e}\right)_{L\oplus N} = \bordermatrix{ & e_{Lj} & L_L^- \cr 
			\overline{e_L}_{i} & (s_W^2-\tfrac{1}{2})\delta_{ij} & 0 \cr 
			\overline{L_L^-} & 0 & s_W^2-\tfrac{1}{2}} \qquad
		&&\left(\mathcal{G}_L^{Z e}\right)_{L\oplus N} = \bordermatrix{ & e_{Rj} & L_R^- \cr 
			\overline{e_R}_{i} & s_W^2\delta_{ij} & 0 \cr 
			\overline{L_R^-} & 0 & s_W^2-\tfrac{1}{2}} \\
		&\left(\mathcal{G}_L^{Z\nu}\right)_{L\oplus N} = \bordermatrix{ & \nu_{Lj} & L_L^0 & N_L \cr 
			\overline{\nu_L}_{i} & \tfrac{1}{2}\delta_{ij} & 0 & 0 \cr 
			\overline{L_L^0} & 0 & \tfrac{1}{2} & 0\cr
			\overline{N_L} & 0 & 0 & 0} 
		&&\left(\mathcal{G}_R^{Z\nu}\right)_{L\oplus N} = \bordermatrix{ & 0 & L_R^0 & N_R \cr 
			0 & 0 & 0 & 0 \cr 
			\overline{L_R^0} & 0 & \tfrac{1}{2} & 0\cr
			\overline{N_R} & 0 & 0 & 0}
	\end{alignat}
\end{subequations}
And finally, the $W$ coupling matrices are obtained from Tab.~\ref{tab:T+} in a similar way,
\begin{align}
	\left(\mathcal{G}_L^{W\nu e}\right)_{L\oplus N} = \bordermatrix{ & e_{Lj} & L^-_L \cr 
				\overline{\nu_{L}}_{i} & \delta_{ij} & 0 \cr 
				\overline{L_L^0} & 0 & 1 \cr
				\overline{N_L} & 0 & 0 } \qquad
	\left(\mathcal{G}_R^{W\nu e}\right)_{L\oplus N} = \bordermatrix{ & e_{Rj} & L^-_R \cr 
				0 & 0 & 0 \cr 
				\overline{L_R^0} & 0 & 1 \cr
				\overline{N_R} & 0 & 0 }.	
\end{align}

\section{One-loop expressions of the form factors}\label{App:FF}

Here we collect the one-loop (mass basis) results specific to the VLL models obtained using \texttt{FeynCalc} \cite{Shtabovenko:2023idz}
and \texttt{FeynArts} \cite{Hahn:2000kx}. In the following, the upper sign correspond to $F_M$ and the lower to $F_D$.

\paragraph{Photon Contribution}
\begin{equation}
	\begin{gathered}
		\includegraphics[width=.2\textwidth]{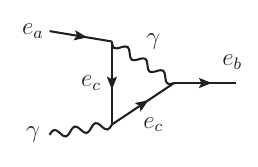} \\
		F_M^{ba}(q^2) = \frac{\alpha}{\pi} \delta_{ac}\delta_{cb}\; m_a^2 [C_1+C_{11}]
		(m_a^2,m_a^2,q^2;m_a^2,0,m_a^2) \qquad F_D^{ba}(q^2)=0
	\end{gathered}
\end{equation}

\paragraph{Higgs contribution}
\begin{equation}
	\begin{gathered}
		\includegraphics[width=.2\textwidth]{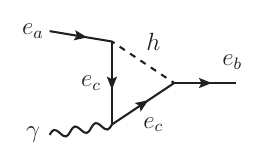} \\
		\begin{aligned}
			F_{M/D}^{ba}(q^2) = \frac{m_a+m_b}{64\pi^2} \bigg\{ &\Big[Y^{h}_{e,ca}Y^{h}_{e,bc} \pm Y^{h*}_{e,ac}Y^{h*}_{e,cb}\Big] \mathcal{F}^\gamma_{h1}
			+ \Big[Y^{h*}_{e,ac}Y^{h}_{e,bc} \pm Y^{h}_{e,ca}Y^{h*}_{e,cb}\Big] \mathcal{F}^{\gamma\pm}_{h2}
			\bigg\}
		\end{aligned}
	\end{gathered}	
\end{equation}
where
\begin{subequations}
	\begin{align}
		\mathcal{F}^\gamma_{h1} &= -m_c \; [C_0+C_1](m_a^2,m_b^2,q^2;m_c^2,M_h^2,m_c^2)\\[.4cm]
		\mathcal{F}^{\gamma\pm}_{h2} &= \big[m_a (C_1+C_{11}) + (m_a \mp m_b) C_{12}\big](m_a^2,m_b^2,q^2;m_c^2,M_h^2,m_c^2).
	\end{align}
\end{subequations}

\paragraph{$Z$ contribution}
\begin{equation}
	\begin{gathered}
		\includegraphics[width=.2\textwidth]{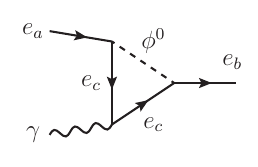} \includegraphics[width=.2\textwidth]{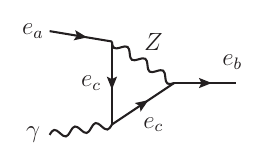} \\
		\begin{aligned}
			F_{M/D}^{ba}(q^2) = \frac{m_a+m_b}{32\pi^2 M_Z^2} \bigg\{ &\Big[g^{Ze}_{R,ca}g^{Ze}_{L,bc} \pm g^{Ze}_{L,ca}g^{Ze}_{R,bc}\Big] \mathcal{F}^{\gamma\pm}_{Z1}
			+ \Big[g^{Ze}_{L,ca}g^{Ze}_{L,bc} \pm  g^{Ze}_{R,ca}g^{Ze}_{R,bc} \Big] \mathcal{F}^{\gamma\pm}_{Z2}
			\bigg\}
		\end{aligned}
	\end{gathered}	
\end{equation}
where
\begin{subequations}
	\begin{align}
		\begin{split}
			\mathcal{F}^{\gamma\pm}_{Z1} &= -m_c \Big[ (m_c^2\pm m_a m_b) C_0 + \big(m_c^2 + m_a^2 - 4M_Z^2 \pm 2m_a m_b\big) C_1 \\
			&\qquad \big[m_a \pm m_b\big] \big(m_a C_{11} + [m_a\mp m_b] C_{12}\big) \Big] (m_a^2,m_b^2,q^2;m_c^2,M_Z^2,m_c^2)
		\end{split} \\[.4cm]
		\begin{split}
			\mathcal{F}^{\gamma\pm}_{Z2} &=\Big[m_c^2(m_a\pm m_b) C_0 + \big[ 2 m_a m_c^2 \pm m_b(m_c^2+m_a^2-2M_Z^2)\big] C_1 \\
			&\qquad + \big[m_c^2 + 2M_Z^2 \pm m_a m_b\big]\big(m_a C_{11} + [m_a\mp m_b] C_{12}\big) \Big] (m_a^2,m_b^2,q^2;m_c^2,M_Z^2,m_c^2).
		\end{split}
	\end{align}
\end{subequations}

\paragraph{$\nu$ contribution}
\begin{equation}
	\begin{gathered}
		\includegraphics[width=.2\textwidth]{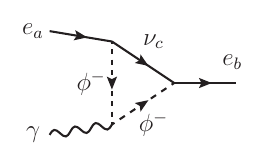} \includegraphics[width=.2\textwidth]{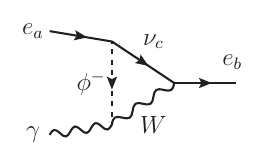}
		\includegraphics[width=.2\textwidth]{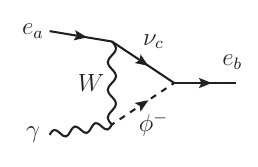} \includegraphics[width=.2\textwidth]{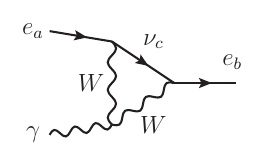} \\
		\begin{aligned}
			F_{M/D}^{ba}(q^2) = \frac{m_a+m_b}{32\pi^2 M_W^2} \bigg\{ &\Big[g^{W\nu e}_{R,ca}g^{W\nu e*}_{L,cb} \pm g^{W\nu e}_{L,ca}g^{W\nu e*}_{R,cb}\Big] 
			\mathcal{F}^{\gamma\pm}_{\nu1}  \\
			+ &\Big[g^{W\nu e}_{L,ca}g^{W\nu e*}_{L,cb} \pm g^{W\nu e}_{R,ca}g^{W\nu e*}_{R,cb} \Big] \mathcal{F}^{\gamma\pm}_{\nu2}
			\bigg\}
		\end{aligned}
	\end{gathered}	
\end{equation}
where
\begin{subequations}
	\begin{align}
		\begin{split}
			\mathcal{F}^{\gamma\pm}_{\nu1} &= m_{\nu_c} \Big[ 4M_W^2 C_0 + \big(m_a^2 - m_{\nu_c}^2 + 4M_W^2\big) C_1 \\
			&\qquad \big[m_a \pm m_b\big] \big(m_a C_{11} + [m_a\mp m_b] C_{12}\big) \Big] (m_a^2,m_b^2,q^2;M_W^2,m_{\nu_c}^2,M_W^2)
		\end{split} \\[.4cm]
		\begin{split}
			\mathcal{F}^{\gamma\pm}_{\nu2} &= - \Big[2M_W^2(m_a\pm m_b) C_0 + \big[ 4 m_a M_W^2 \pm m_b (m_a^2 - m_{\nu_c}^2 + 2M_W^2)\big] C_1 \\
			&\qquad + \big[m_{\nu_c}^2 + 2M_Z^2 \pm m_a m_b\big]\big(m_a C_{11} + [m_a\mp m_b] C_{12}\big) \Big] (m_a^2,m_b^2,q^2;M_W^2,m_{\nu_c}^2,M_W^2).
		\end{split}
	\end{align}
\end{subequations}
\paragraph{$\rho$ contribution}
\begin{equation}
	\begin{gathered}
		\includegraphics[width=.2\textwidth]{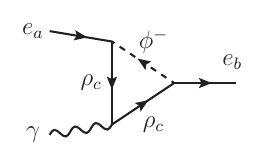} \includegraphics[width=.2\textwidth]{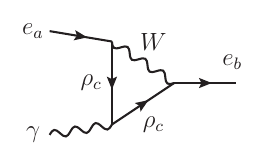} \\
		\begin{aligned}
			F_{M/D}^{ba}(q^2) = \frac{m_a+m_b}{16\pi^2 M_W^2} \bigg\{ &\Big[g^{We\rho*}_{R,ac}g^{We\rho}_{L,bc} \pm 
			g^{We\rho*}_{L,ac}g^{We\rho}_{R,bc}\Big] \mathcal{F}^{\gamma\pm}_{W1} \\
			+ &\Big[g^{We\rho*}_{L,ac}g^{We\rho}_{L,bc} \pm  g^{We\rho*}_{R,ac}g^{We\rho}_{R,bc} \Big] \mathcal{F}^{\gamma\pm}_{W2}
			\bigg\}
		\end{aligned}
	\end{gathered}	
\end{equation}
where $\mathcal{F}^{\gamma\pm}_{Wi}$ are equal to $\mathcal{F}^{\gamma\pm}_{Zi}$ with $M_Z\to M_W$ and $m_c \to m_{\rho_c}$.
\begin{equation}
	\begin{gathered}
		\includegraphics[width=.2\textwidth]{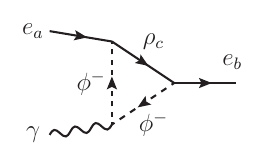} \includegraphics[width=.2\textwidth]{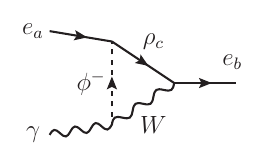}
		\includegraphics[width=.2\textwidth]{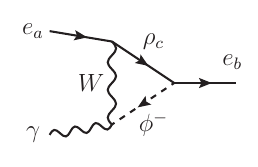} \includegraphics[width=.2\textwidth]{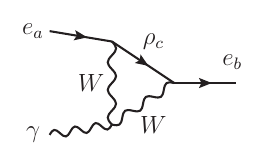} \\
		\begin{aligned}
			F_{M/D}^{ba}(q^2) = -\frac{m_a+m_b}{32\pi^2 M_W^2} \bigg\{ &\Big[g^{We\rho*}_{R,ac}g^{We\rho}_{L,bc} \pm 
			g^{We\rho*}_{L,ac}g^{We\rho}_{R,bc}\Big] 
			\mathcal{F}^{\gamma\pm}_{\rho1}\\
			+ &\Big[g^{We\rho*}_{L,ac}g^{We\rho}_{L,bc} \pm g^{We\rho*}_{R,ac}g^{We\rho}_{R,bc} \Big] \mathcal{F}^{\gamma\pm}_{\rho2}
			\bigg\}
		\end{aligned}
	\end{gathered}	
\end{equation}
where $\mathcal{F}^{\gamma\pm}_{\rho i}$ are equal to $\mathcal{F}^{\gamma\pm}_{\nu i}$ with $m_c \to m_{\rho_c}$.

\printbibliography
	
\end{document}